\documentclass[12pt,preprint]{aastex}

\shorttitle{Dense Molecular Clumps in the LMC}
\shortauthors{Seale et al.}

\begin{document}

\title{The Life and Death of Dense Molecular Clumps in the Large Magellanic Cloud}

\author{}
\affil{}

\author{Jonathan P. Seale\altaffilmark{1}, Leslie W. Looney, Tony Wong}
\affil{Astronomy Department, University of Illinois \\ Department of Astronomy, MC-221 \\ 1002 W. Green Street \\ Urbana, IL 61801}

\author{J$\rm\ddot{u}$rgen Ott}
\affil{National Radio Astronomy Observatory \\ P.O. Box O \\ 1003 Lopezville Road \\ Socorro, NM 87801}

\author{Uli Klein}
\affil{Universit$\ddot{a}$t Bonn \\ Argelander-Institut f$\ddot{u}$r Astronomie \\ Auf dem H$\ddot{u}$gel 71 \\ D-53121 Bonn, Germany }

\author{Jorge L. Pineda}
\affil{Jet Propulsion Laboratory, California Institute of Technology \\ 4800 Oak Grove Drive \\ Pasadena, CA}

\altaffiltext{1}{current address: Space Telescope Science Institute, 3700 San Martin Dr., Baltimore, MD 21218}

\begin{abstract}

We report the results of a high spatial (parsec) resolution HCO$^{+}$ ($J = 1\to0$) and HCN ($J = 1\to0$) emission survey toward the giant molecular clouds of the star formation regions N\,105, N\,113, N\,159, and N\,44 in the Large Magellanic Cloud. The HCO$^{+}$ and HCN  observations at 89.2 and 88.6 GHz, respectively, were conducted in the compact configuration of the Australia Telescope Compact Array. The emission is imaged into individual clumps with masses between $10^{2}$ and $10^{4}$ M$_{\sun}$ and radii of $<1$ pc to $\sim2$ pc. Many of the clumps are coincident with indicators of current massive star formation, indicating that many of the clumps are associated with deeply-embedded forming stars and star clusters. We find that massive YSO-bearing clumps tend to be larger ($\gtrsim$1 pc), more massive (M $\gtrsim 10^{3}$ M$_{\sun}$), and have higher surface densities ($\sim1$ g cm$^{-2}$), while clumps without signs of star formation are smaller ($\lesssim$1 pc), less massive (M $\lesssim 10^{3}$ M$_{\sun}$), and have lower surface densities ($\sim0.1$ g cm$^{-2}$). The dearth of massive (M $>10^{3}$ M$_{\sun}$) clumps not bearing massive YSOs suggests the onset of star formation occurs rapidly once the clump has attained physical properties favorable to massive star formation. Using a large sample of LMC massive YSO mid-IR spectra, we estimate that $\sim2/3$ of the massive YSOs for which there are \textit{Spitzer} mid-IR spectra are no longer located in molecular clumps; we estimate that these young stars/clusters have destroyed their natal clumps on a time scale of at least $\sim3\times10^{5}$ yrs.

\end{abstract}


\section{Introduction}

Despite their dominant role in shaping galactic structures and stellar content, the current understanding of massive star formation (M $\gtrsim8$ M$_{\sun}$ ) remains incomplete. There are a number of reasons for this, many related to observational difficulties. Massive stars form deeply embedded in the densest regions of molecular clouds with high extinctions, making them difficult to observe during their early formation stages \citep[e.g.,][]{lada03}. Due to the nature of the stellar initial mass function \citep[IMF;][]{salp55}, they are rare objects, and coupled with their short formation times \citep[$\sim10^{5}$ yrs; e.g., see][and references therein]{chur02}, identifying observational targets is difficult. Furthermore, high-mass stars are seldom found in isolation \citep[e.g.,][]{prei01}, meaning the interpretation of observations is complicated by the presence of other cluster members, the local environment, outflows, winds, and ionizing radiation.

In recent years a rough evolutionary scenario for massive star formation has emerged \citep[see recent reviews by][]{beut07, mcke07, zinn07}. Stars are born within giant molecular clouds (GMCs), which have radii on the order of tens of parsecs and masses that can exceed $10^{4}$ M$_{\sun}$. These GMCs are hierarchical in nature, with substructures that span a range of scales, and the literature contains a confused lexicon of terminology describing this substructure. Following the nomenclature of \citet{will00}, the largest (approximately parsec-sized) over-dense regions with typical masses of $10^{2} - 10^{3}$ M$_{\sun}$ are defined as \textit{clumps}. A \textit{core} is a smaller, denser region of molecular gas that will form a single star or a small multiple system. The details of how massive star formation proceeds from these basic units within the GMC is unclear. A protostar forms from gravitational collapse within the core, and it may grow in mass by accretion of the immediate core material \citep[e.g.,][]{mcke03}. Alternatively, the star may grow from the accretion of clump material initially unbound from the core \citep[e.g.,][]{bonn97}, with the most massive members of a cluster forming near the center of the clump where the gravitational potential and resulting accretion rates are highest. The Kelvin-Helmholtz contraction timescale of a high-mass star is smaller than the star's accretion timescale, meaning accretion can still be occurring after the star has begun hydrogen burning. The radiation of the star will ionize its surrounding to an increasingly larger scale until a combination of radiation pressure, stellar winds, and outflows have dissipated the dense surrounding material. In this paper, we observationally explore the properties of clumps as they evolve from dense star-forming structures to the less dense medium dispersed by the newly-formed stars.

Observations of dust and molecules can trace the core and clump material being converted into stars and provide direct insight into the initial conditions of star formation. There have been numerous quantitative studies of the properties of these GMC substructures conducted in Milky Way star formation regions \citep[e.g.,][]{stut88,will94,elme96,kram98}. These gas and dust surveys have revealed the dense clumps to have densities of n$_{\rm{H_{2}}}\approx10^{5}$cm$^{-3}$, gas temperatures of $10-30$ K, radii of 0.1--1 pc, and gas masses ranging from a few hundred to tens of thousdands of M$_{\sun}$ that are found to be consistent with a power law dN/dM$\propto$M$^{\alpha}$, where $\alpha$ is typically found to be between $-1$ and $-2$. The mass spectrum extends to lower-mass overdensities essentially unbroken over many orders of magnitude. Kinematic information about the dense gas is accessible through the rotational transitions of molecules with large dipole moments(e.g., CS, HCN, HCO$^{+}$), and the massive clumps are found to be gravitationally bound structures.

Given the spatial resolution limitations, the properties of the dense gas in extragalactic star formation regions is less constrained than in the Galaxy. Using the Swedish-ESO 15m Submillimeter Telescope (SEST), \citet{joha94} report the detection of eight molecules ($^{12}$CO, CS, SO, CN, HCN, HNC, SCO$^{+}$, and H$_{2}$CO) towards the LMC's N\,159 HII complex, and \citet{chin97}, who extended the observations to the N\,113, N\,44, and N\,214 complexes, detected $^{12}$CO, $^{13}$CO, CN, CS, HNC, HNC, HCO$^{+}$, HC$_{2}$, and C$_{3}$H$_{2}$. While the gas was unresolved in SEST's large beam ($\sim50\arcsec$), the presence of dense gas is implied by the detection of the emission lines of molecules with high critical densities. With multiple transitions, it is possible to estimate the gas temperature and density, and from $^{12}$CO and $^{13}$CO observations, \citet{mina11} finds a temperature of $\rm{T}=15-200$K and density of $\rm{n}_{\rm{H}_{2}}=10^{3}$cm$^{-3}$ in 32 $^{13}$CO clumps in 6 LMC GMCs. \citet{bola00} finds a dense gas emission component of T=$15$ K and $\rm{n}_{\rm{H}_{2}}=10^{5}$cm$^{-3}$ from $^{12}$CO and $^{13}$CO observations of N159 by using a two-phase gas model that also included a T=$100$ K and $\rm{n}_{\rm{H}_{2}}=10^{2}$cm$^{-3}$ phase. 

Such low spatial resolution observations guided several targeted observations of single clumps in order to more robustly determine their properties. Determining the total mass contained within the cold, dense phase implied by the above molecular detections is only possible if the dense gas is spatially resolved. Higher resolution interferrometric observations by\citet{ott10} of one of the emission peaks in N\,159 resulted in the first detection of NH$_{3}$ in the LMC; the two detected lines implied gas temperature of $\sim16$ K and a total gas mass of $10^{4}$M$_{\sun}$. \citet{wong06} resolved a clump in N\,113 with targeted HCN and HCO$^{+}$ observations, determining a clump virial mass of $\sim10^{4}$. Here we report the results of a high resolution study of dense molecular clumps in several star formation regions in the Large Magellanic Cloud (LMC) using HCO$^{+}$ ($J = 1\to0$) and HCN ($J = 1\to0$) as high-density gas tracers. The observations, the first systematic interferometric mapping of dense gas in the LMC, were conducted with the Australia Telescope Compact Array (ATCA).

Some of the problems of studying star formation in the Galaxy can be mitigated in the LMC. Galactic star formation is concentrated along the plane of the Galaxy, where dust obscuration is high; however, the LMC is located at high Galactic latitude with little intervening Galactic material. Milky Way studies are also plagued by distance uncertainties, while the LMC is located at a known distance of 50 kpc \citep[for a review of recent distance calculations, see][]{alve04}, making luminosity determinations relatively robust. One of the primary observational advantages to studying the LMC is its proximity relative to other extragalactic systems, which allows single massive YSOs (or tightly bound systems) to be resolved even in the mid-IR. Taking advantage of the ability to observe the entire galaxy, a number of studies have sought to identify and catalog all the massive and intermediate-mass young stellar objects (YSOs) in the LMC \citep{whit08,grue09}. Combined with other studies that identified sites of on-going star formation via their infrared emission, maser activity, and compact H$\alpha$ emission and centimeter emission, the literature now contains an extensive catalog of LMC YSOs \citep[e.g.,][]{inde04,chen09,chen10,elli10} to which we will positionally compare the dense gas revealed by our observations in order to explore the relationship between star formation and the dense gas.

We have conducted HCO$^{+}$ ($J = 1\to0$) and HCN ($J = 1\to0$) observations with the ATCA of four active star formation regions in the LMC -- N\,105, N\,113, N\,159, and N\,44. With critical densities of $>10^{5}$ cm$^{-3}$ at T$=$20K \citep{scho05}, HCO$^{+}$ and HCN were selected to probe the high volume densities of molecular clumps. We have analyzed the observations to determine the properties of the dense gas and its association with ongoing star formation in the LMC. Section 2 describes our observations and data reduction procedure. Section 3 details how discrete regions of molecular emission are identified, how their properties are determined, and the distributions of these clump properties. In Section 4 we summarize the star formation activity within the four complexes and explore its relationship with the imaged molecular material. Finally, a discussion of the feedback from star formation into the clumps is presented in Section 5; included is a scenario relating the properties of the clumps to the evolutionary state of the young stellar population, which we use to derive the timescales involved in the massive star formation process.

\section{Observations}

The regions chosen for observation in this study were selected for their known association with massive star formation. N\,105, N\,113, N\,159, and N\,44 are well-studied \ion{H}{2} complexes in the LMC, each containing one to several GMCs with signatures of both recent (H$\alpha$ emission) and on-going (e.g., maser and bright IR emission) star formation. For each complex, we targeted the $^{12}$CO emission peak of the region's GMC(s) with the aim of imaging the dense gas associated with current star formation in the cloud. Figures 1--5 show recent single-dish $^{12}$CO observations from the Magellanic Mopra Assessment \citep[MAGMA;][]{hugh10,wong11}, a survey of the molecular content of the Large and Small Magellanic Clouds using the Mopra Telescope along with the 50$\%$ sensitivity of our ATCA observations toward the CO emission peaks. Note that with a primary beam full width at half maximum (FWHM) of $\sim33\arcsec$, to cover large sections of the molecular clouds, mosaics were composed consisting of multiple ATCA pointings.

The 3 mm HCO$^{+}$ and HCN data were collected during 12 separate sets of observations between Septermber 2006 and September 2009. At the time, the ATCA correlator could be configured to observe two frequencies simultaneously in dual polarization mode, allowing us to observe both HCO$^{+}$ ($J = 1\to0$; 89.1885 GHz rest frequency) and HCN ($J = 1\to0$; 88.6318 GHz rest frequency), redshifted to the appropriate observational frequency. All observations were conducted using the same array configuration (H75), which has a maximum antenna spacing of approximately 90 m, giving a synthesized beam FWHM of approximately 6$\arcsec$. Table 1 lists the sky position of each region, dates of the observations, number of pointings in the mosaic, HCO$^{+}$ synthesized beam sizes, typical achieved sensitivities (similar for both frequencies), and frequency channel width of the observations. Note that we have separated N\,44 into two distinct, non-contiguous regions: N\,44's northern molecular cloud, which we refer to as N\,44 Region 1, and the southernmost molecular cloud, N\,44 Region 2.

The data were calibrated and images created using the Australia Telescope National Facility (ATNF) version of the \textsf{MIRIAD} package \citep{saul95}. Flux calibration was performed using Uranus, PKS 1921--293, and Mars for N\,105, N\,113, and N\,44, respectively; N\,159 was flux calibrated using PKS 0537--752, PKS 1253-055, and Mars over its 5-day observation. The overall flux calibration uncertainty is estimated  at $20\%$. Quasar PKS 0537--752 was used as the phase calibrator for all observations. PKS 1921--293 was used to calibrate the bandpass for N\,105 and N\,113; PKS 1253-055 was used for N\,159 and N\,44. After imaging the maps were CLEANed to a 2-$\sigma$ level. These cleaned data cubes are then used in the clump-identification procedure described below. The contours and gray scale pixels in the bottom panels of Figures 1--5 are $0^{\rm{th}}$ moment (integrated intensity) maps of the HCO$^{+}$ and HCN data cubes, masked with a smooth version to reduce noise.

To test the effect of the observations' uv coverage on the recovered source flux, we modeled an observation of a Gaussian source using the mosaic pattern and uv-coverage of the N105 mosaic. The model source was placed near the center of the mosaic, and fake visibilities were generated and imaged using the same procedure outlined above for the actual data. We varied the full-width-half-maximum (FWHM) of the model source to be between 3$\arcsec$ and 21$\arcsec$; $> 80\%$ of the flux is recovered for sources with FWHM$<9\arcsec$ and fluxes can be measured within a  factor of 2 accuracy for intrinsic sizes of FWHM=$14\arcsec$ or smaller. Larger sources tend to be resolved out. A poor signal to noise ratio reduces the recovered flux fraction roughly similarly for all size scales. By the definition of source radius defined in the following section, we are sensitive to structures of radius R$\leq 2.6$ pc at the distance of the LMC.

\section{Identification and Characterization of Dense Molecular Clumps}

The HCO$^{+}$ and HCN observations show that the molecular material within N\,105, N\,113, N\,159, and N\,44 is clumpy with emitting clouds of varied sizes, shapes, and velocity structures (Figures 1--5). HCO$^{+}$ and HCN had previously been detected in N\,113's GMC with lower angular resolution single dish observations with the SEST \citep{joha94,chin97,heik99} and Mopra telescopes \citep{wong06}. However, the implied density of the GMC is $\sim200$ cm$^{-3}$ \citep{wong06}, significantly lower than the critical densities of HCO$^{+}$ ($1-0$) and HCN ($1-0$) \citep[$1.8 \times 10^{5}$ and $1.2 \times 10^{6}$ cm$^{-3}$, respectively, at 20K; from the Leiden Atomic and Molecular Database,][]{scho05}. The implication is that the molecular cloud is clumped into substructures, and our mosaic ATCA observations directly reveal this clumpiness. We investigate the variations in clump properties quantitatively by determining the sizes, linewidths, and masses of the clumps.

\subsection{Identifying Clumps}

We use the HCO$^{+}$ ($J = 1\to0$) data as the primary clump tracer. The HCN data has a lower overall signal-to-noise ratio (SNR) than that of HCO$^{+}$ because of the typically weaker HCN flux. To identify HCO$^{+}$ clumps in the ATCA data we use the algorithm described in \citet{roso06} by using the software version of the procedure available as an \textsf{IDL} package called \textsf{CPROPS}. The procedure is described in detail by \citet{roso06} and in the \textsf{CPROPS} users guide, but a description of the basic steps of the algorithm follows. A mask is first generated from the data to isolate significant emission by estimating the noise variance ($\sigma_{\rm{rms}}$) and identifying emission cores of four adjacent channels with flux density $\ge2.5\sigma_{\rm{rms}}$. Each core is then expanded to include all connected emission above the $2\sigma_{\rm{rms}}$ level. These masking levels (called the `threshold' and `edge' values in \textsf{CPROPS}) can be user-defined, and we explored a range of values. We find that threshold $= 2.5\sigma_{rms}$ and edge $= 2\sigma_{\rm{rms}}$ were able to distinguish significant emission from noise while also identifying faint clumps.

Once the images have been broken into these `islands' of significant emission, \textsf{CPROPS} assigns the emission to individual clouds. Molecular clouds are known to be clumpy, and the islands identified here often contain substructure themselves. When a single surface brightness contour contains multiple emission peaks, \textsf{CPROPS} attempts to separate the island into multiple clumps. The peaks are separated into distinct clumps if each clump is larger than 1 resolution element (the synthesized beam), the contrast between each peak and edge of the clump exceeds a threshold, and the individual clumps have significantly different measured properties (e.g., mass, velocity width) compared to the combined island. When separating the islands into individual clumps with \textsf{CPROPS}, we chose to use the \textsf{ECLUMP} keyword that allows emission shared within a single brightness contour level by two separate clumps to be partitioned between the clumps. The \textsf{ECLUMP} keyword invokes a `friends-of-friends' algorithm such that the pixels immediately surrounding a clump (friends) are incorporated into that particular clump, pixels immediately surrounding these friend pixels (friends of friends) are then incorporated, and so on, until all pixels are assigned to a clump. Pixels that are `friends' with multiple clumps are assigned to the clump of nearest projected distance. \textsf{CPROPS} identified a total of 46 HCO$^{+}$ clumps in the five regions -- five in N\,105, six in N\,113, twenty in N\,159, nine in N\,44's Region 1, and six in N\,44's Region 2.

Because of the signal-to-noise ratio (SNR) threshold invoked to isolate significant emission, the identification of clumps is sensitive to the depth of the observations. Requiring a clump emission peak to have at least 4 consecutive channels above $2\sigma_{\rm{rms}}$ requires all clumps with broad velocity dispersions ($\sigma_{\rm{v}}$ $\gg$ channel width) to have a peak SNR of more than $\sim3$ within a single velocity channel. Clumps with narrow velocity widths (i.e., with velocity dispersions comparable to or smaller than the spectral resolution) require a higher peak SNR in order to have 4 consecutive channels above the $2\sigma_{\rm{rms}}$ threshold. For a clump with a narrow velocity FWHM of 1 km s$^{-1}$, observations with channel widths of 0.2 and 0.4 km s$^{-1}$ would require clump core detections of 3.2$\sigma_{\rm{rms}}$ and 3.9$\sigma_{\rm{rms}}$ in a spectral channel, respectively. These detection thresholds amount to a slightly different minimum detectable flux limit for each region. The minimum detectable HCO$^{+}$ flux for an unresolved clump with a velocity FWHM of 1 km s$^{-1}$ is $F_{HCO^{+}}=0.19$ Jy km s$^{-1}$ in N\,105, whose observations have the lowest noise. On the other hand, N\,113 and N\,44 Region 2 have the highest minimum detectable flux of $F_{HCO^{+}}=0.26$ Jy km s$^{-1}$. These detection thresholds are very similar and lower than all save one clump's flux, and we therefore assert that despite slightly different noise levels and channel widths, the observations in each region are sensitive to the same clump population.

While most of the HCO$^{+}$ emission peaks are strong detections with a high spectral SNR (up to SNR$\approx10$), those with peak SNR's of $\sim3$ may be described as weak and/or possible non-detections. Although our threshold requirement of 4 consecutive channels above $2\sigma_{\rm{rms}}$ generally does a good job of separating noise from true sources, we allow for the possibilty of false detections by labeling weakly-emitting peaks as `candidate clumps;' their classification as real clumps will require deeper imaging. Candidate clumps, indicated by the ``c'' supercript in Table 2, are identified by narrow line widths (which could be mimicked by noise) and a low spectral SNR at clump center and across the entire clump (Figure 6). We assert all other detections are strong enough to indicate the emission is real and originates from true dense clumps.

\subsection{Derivation of Clump Physical Properties}

The radii of the clumps are determined by \textsf{CPROPS}, and in Table 2, we list the results. A clump's major and minor axes are first identified using principal component analysis (PCA; identifies the axis along which there is the most variation), and the second moments along each of these axes is then determined. The rms radius is the geometric mean of these two second moments. This measured size is necessarily an underestimate due to the sensitivity limits of the observation; any flux belonging to the clump below the detection threshold will not be used in the above calculation. To account for this missing flux, \textsf{CPROPS} measures the size of the clump as a function of minimum surface brightness (essentially varying the observation's noise) and extrapolates the size of the clump to a value with zero noise. This method can also help mitigate the problem of varying sensitivity limits between different observations. The synthesized beam is deconvolved from the resulting rms size, which is then converted to clump radius by assuming a particular brightness profile and distance. If the extrapolated clump size from the cleaned image is smaller than the synthesized beam, it cannot be deconvolved. Furthermore, marginally-resolved clumps suffer from substantial radii uncertainties. As can be seen in Figures 1-5, many clumps are either unresolved or are comparable in size to the beam ($\sim1-1.5$ pc at the distance of the LMC). This is consistent with the sizes of dense molecular clumps in Galactic massive star forming regions with measured sizes of $\sim0.1 - 1$ pc. Given the angular resolution limitations of extragalactic 3 mm observations, at the distance of the LMC, only the largest ($\gtrsim1$ pc) clumps are resolved.

A visual inspection of the ATCA data shows that the HCO$^{+}$ emission for each clump emanates from a somewhat larger region than the HCN emission. \citet{wong06} estimates the HCN-determined radius of N\,113's clump 4 is $89\%$ of that determined from HCO$^{+}$ and finds a decreasing HCO$^{+}$/HCN ratio on longer interferometer baselines. As larger baselines probe smaller scales, both findings suggests that the HCO$^{+}$ emission originates from a larger region than HCN. Such a radial dependence could be explained if HCN probes higher densities than HCO$^{+}$. Indeed, HCN ($J = 1\to0$) has a critical density of $n_{\rm{crit}}$ $=$ 1.2 $\times$ 10$^{6}$ at 20 K, compared to $n_{\rm{crit}}$ $=$ 1.8 $\times$ 10$^{5}$ for HCO$^{+}$ ($J = 1\to0$) \citep{scho05}. Moreover, the HCO$^{+}$ abundance may be reduced in the high density centers of clumps due to its participation in proton transfer reactions with abundant neutral molecules \citep[][]{heat93}, and the abundance of HCO$^{+}$ may be enhanced in the outer, photon-dominated clump layers via chemical reactions involving C$^{+}$, which is expected to be abundant in regions exposed to UV radiation \citep[][]{heik99}. Each of the \ion{H}{2} complexes studied here contain a number of UV-producing O and B-type stars, plausible sources of ionizing radiation that could enhance the HCO$^{+}$ abundance in the outer parts of the clump. In the latter case, the size determined from HCO+ in a high ionizing radiation field region is larger than would be found with less ionizing radiation.

\textsf{CPROPS} calculates the velocity dispersion $\sigma_{\rm{v}}$ (the velocity size) of each clump from the second moment along the velocity axis of all clump pixels. Like the rms clump size, the velocity dispersion is extrapolated to what would be expected from a zero-noise observation and deconvolved with the channel width. For a Gaussian line profile, the full-width half-maximum (FWHM, $\Delta$v) line width is then given by $\Delta$v $=\sqrt{ 8\ \rm{ln}(2)}\ \sigma_{\rm{v}}$ and is presented in Table 2.

Numerous previous studies of massive star forming clumps indicate that the masses of the clumps can be estimated via the virial theorem \citep[e.g.,][]{lars81,solo87,sait06,wong06,mull10}. Assuming the clumps are spherical, have radial density profiles described by a power law, and are virialized, the virial mass is given by the formula \citep{solo87}

\begin{equation}
\rm{M}_{\rm{vir}} = 125 \rm{M}_{\sun}\frac{ (5-2\beta)}{(3-\beta)}\ \Delta\rm{v}^{2}\rm{R},
\end{equation}

\noindent where $\beta$ is the power law exponent of the density profile (density $\propto\ r^{-\beta}$; r is the distance from clump center), R is the clump radius in pc, and $\Delta$v is the FWHM of the velocity line width in km sec$^{-1}$. The virial masses provided in Table 2 assume the \textsf{CPROPS}-determined radius and a density profile with $\beta=1$ \citep{vandertak00}; the provided uncertainty assumes only observational effects (uncertainties in $\Delta$v and R) and does not account for the uncertainty of $\beta$. Clumps with extrapolated sizes smaller than the beam have no radius estimates, and therefore M$_{\rm{vir}}$ cannot be determined.

The masses of the clumps can also theoretically be determined directly from the HCO$^{+}$ line fluxes if the HCO$^{+}$ abundance and the optical depth of the line can be estimated, which requires the excitation temperature T$_{\rm{ex}}$) to be known. Because we only have one transition of HCO$^{+}$, it is not possible to determine T$_{\rm{ex}}$ from our data, so in order to estimate the masses of the clumps from the HCO$^{+}$ flux, we assume a single value of T$_{\rm{ex}}$=30K for all clumps, typical of massive clumps observed in dust continuum \citep[e.g.,][]{faun04,belt06}. We will investigate the effect of choosing a different value for T$_{\rm{ex}}$ below. We estimate $\tau$, the clump's optical depth, from $\tau$ = $-$ln[1 -- T$_{b}$/(T$_{\rm{ex}}$ -- T$_{\rm{bg}}$)], where T$_{b}$ is the clump's typical brightness temperature, taken to be 1/2 the clump's peak brightness temperature \citep{barn11};  T$_{\rm{ex}}$ is the excitation temperature; and  T$_{\rm{bg}}$ is the background temperature (2.72 K). We can then estimate the clumps' masses under LTE by calculating the column density and integrating over the surface area of the clump. Following the formulation of \citet{barn11}, 

\begin{equation}
\rm{M}_{\rm{LTE}} = 4.0 \times 10^{3}\, \rm{M}_{\sun}\, \Omega\, \int \tau\, \rm{dV}
\end{equation}

\noindent where V, the line velocity, is in km s$^{-1}$ and $\Omega$, a clump's projected surface area, is in pc$^{2}$. Because we are assuming an excitation temperature and HCO$^{+}$ abundance \citep[n(HCO$^{+}$)/n(H$_{2}$) = 10$^{-9}$; the upper limit of recent core chemistry models and a typical value derived for massive clumps; see][and references therein]{barn11}, M$_{\rm{LTE}}$ is only an estimate. Resolved clumps have M$_{\rm{LTE}}$ uncertainties of typically $\sim40\%$ derived from the uncertainty in clump size and velocity dipersion. Because calculating M$_{\rm{LTE}}$ requires T$_{b}$ and $\Omega$, its estimation for unresolved clumps is more uncertain. For the unresolved clumps, Table 2 reports M$_{\rm{LTE}}$ by assuming the clump is the size of the beam; the uncertainty of M$_{\rm{LTE}}$ for unresolved clumps is $\sim200\%$, and because the clumps' brightness temperatures suffer from beam dilution, our estimate is likely an underestimate.

The M$_{\rm{LTE}}$ uncertainties in Table 2 take into account the uncertainty of $\Omega$, T$_{b}$, and $\Delta\rm{v}$, but does not consider the uncertaity in $\tau$ resulting from T$_{\rm{ex}}$. T$_{\rm{ex}}$ is not measurable with our data and is potentially the largest source of uncertainty in our M$_{\rm{LTE}}$ estimate. Clumps that truly have a higher excitation (possibly such as those with embedded YSOs) have a lower M$_{\rm{LTE}}$, while in low excitation clumps (e.g., those without embedded YSOs), M$_{\rm{LTE}}$ is higher. The optical depths (and therefore M$_{\rm{LTE}}$) at the lower T$_{\rm{ex}}$ of 10K are at least a factor of 3 higher than at 30K (and higher for the brightest clumps) and a factor of 2 lower at the higher T$_{\rm{ex}}$ of 50K. The analysis presented here is with a uniform T$_{\rm{ex}}$ for uniformity and simplicity.

\subsection{Distribution of Properties}

Table 2 documents the derived properties of the clumps including position, $\Delta$v, radius, M$_{\rm{vir}}$, M$_{\rm{LTE}}$, and total HCO$^{+}$ and HCN flux within the clump. Figure 6 shows the $F_{\rm{HCO^{+}}}$, $F_{\rm{HCN}}$, and M$_{\rm{LTE}}$ distributions for the entire clump sample. In each panel, the histogram displays a binned representation of the distribution whose number counts are indicated by the left axis, while the points are associated with the right axis and indicate each distribution's corresponding cumulative function, which is independent of bin choices. The vertical dotted line shows the location of the sensitivity limit in N\,113 and N\,44 for a clump with size equal to that of the synthesized beam and $\Delta$v = 1 km s$^{-1}$; this corresponds to a clump with M$_{\rm{LTE}}$ of $\sim250$M$_{\sun}$.

The mass function of a population of clouds is typically described by a power law fit to the differential mass distribution, dN/dM$\ \propto\ $M$^{\alpha}$, where $\alpha$ is the power law index of the distribution. N (M$^{\prime}$$>$M), the integral of the mass function above mass M, can then be described by N(M$^{\prime}>$M)$\ \propto\ $M$^{\alpha+1}$. The power law index of the mass function for Galactic cores (0.5 M$_{\sun}$ $<$ M $<$ 20 M$_{\sun}$) has been found to vary between $-1.6$ to $-2.8$ with a typical value of $\sim2$ in local star formation regions such as Perseus \citep{enoc06}, Ophiuchus \citep{youn06}, and Serpens \citep{enoc07}. Recent early results from the \textit{Planck} satellite found a Galactic core and clump (10 M$_{\sun}$ $<$ M $<$ $10^{3}$ M$_{\sun}$) mass distribution consistent with $\alpha=-2$ \citep{plan11}. We find the clump mass distribution to be consistent with $\alpha=-2$, but the uncertainty of M$_{\rm{LTE}}$ prevents a more detailed fitting.

\section{Star Formation Activity within the GMCs}

\subsection{Identifying and Characterizing Star Formation Activity within the GMCs}

Recently, two separate, extensive surveys of LMC star formation were conducted using the \textit{Spitzer Space Telescope} with the aim of identifying the galaxy's YSO population. We use the term YSO to describe any young star (typical ages of several $10^{5}-10^{6}$ years) that is still surrounded by enough circumstellar material to have a substantial infrared excess. Together, these two surveys, \citet[][hereafter W08]{whit08} and \citet[][hereafter GC09]{grue09} identified $1000+$ sources described as candidate YSOs (or tightly-bound, young multiple systems). We have used these two surveys as the primary catalogs from which we determine the locations of YSOs within the \ion{H}{2} complexes studied, but we have also combed the literature for further YSO identifications. In particular, detailed IR studies of N\,44 by \citet{chen09} and of N\,159 by \citet{jone05} and \citet{chen10} have identified several YSOs not identified in GC09 or W08, and numerous compact \ion{H}{2} region \citep[e.g.,][]{inde04,mart05} and maser surveys \citep[e.g.,][]{broo97,laze02,elli10} have been used to strengthen YSO identifications. Figures 1--5 indicate the locations of sources identified as YSOs in the literature and Table 3 lists their positions and properties. The aforementioned means of detecting YSOs are either flux limited or restricted to phenomena only associated with the most massive YSOs. Therefore only massive and intermediate-mass YSOs are marked in the figures.

Based upon SED fits to radiation transfer models of YSO systems, GC09 and \citet{chen09} suggest that the 8.0 $\mu$m flux of a YSO may be a good proxy for the central source's total luminosity, and therefore, its mass. Table 3 lists each YSO's 8.0 $\mu$m magnitude; we classify those with [8.0] $\leq$ 8.0 as massive YSOs, while those with [8.0] $>$ 8.0 are intermediate-mass. GC09 demonstrated that their candidate YSO catalog was nearly complete ($>99\%$) in the explored color-magnitude space for sources brighter than the 8.0 magnitude threshold; we assume all massive YSOs have been identified within the imaged regions.

For the purposes of this paper, the evoluionary state of the complexes' YSO population is of particular interest. Many of the YSOs, particularly the most luminous, have been confirmed spectroscopically by \citet[][hereafter SL09]{seal09} with mid-IR \textit{Spitzer} IRS spectra. SL09 classified the spectra based upon the dominant spectral features and purports these spectral groups represent the evolutionary development of a YSO from one highly embedded in its natal molecular material to more advanced-stage YSOs with emission typical of photodissociation and compact \ion{H}{2} regions. The strength of the silicate absopriton feature may be among the more evolutionarily diagnostic spectral features, and the feature's strength along with each YSO's spectral group is reproduced in Table 2. Note that the silicate strengths were determined using principal component analysis (PCA) and should not be interpreted as the optical depth (nor is it necessarily proportional to the optical depth) of the feature as the spectra are first normalized to a common luminosity (see SL09). Instead, it should be interpreted as a dimensionless measure of the depth of the silicate absorption relative to the continuum that can range between -1 (little to no silicate absorption) to 1 (strong silicate absorption). Moreover, in the detailed analyses of the star formation activity within N\,44 and N\,159, \citet[][]{chen09,chen10}  supplemented \textit{Spitzer} data with optical and near-IR images to construct SEDs from $\sim0.3$ $\mu$m to $\sim100$ $\mu$m. They proposed a classification scheme for YSOs based upon their SED shapes. It is suggested that these Types are evolutionary (Types I, II, and III are arranged youngest to oldest) and reflect the amount of obscuring circumstellar material around the forming star. For the YSOs in N\,44 and N\,159 that were assigned Types by \citet[][]{chen09, chen10}, we have listed their Types in Table 3.

\subsection{Physical Association Between Clumps and Star Formation}

In this study we have identified a total of 36 dense molecular clumps (and 10 candidate clumps) from their HCO$^{+}$ emission. Even a by eye inspection of Figures 1-5 reveals a strong correlation between the location of the YSOs and the clumps' dense gas. The relationship between the YSOs and clumps can be quantified by comparing the proximity of the clumps to the YSOs and the clumps to a hypothetical field of YSOs randomly distributed throughout the region. An object randomly placed in region N\,159 (the region we have most completely imaged) has only a 10$\%$ probability of being located within 12'' of an emission peak of a clump (i.e., only 10$\%$ of N\,159's imaged surface area is within 12$\arcsec$ of a clump's emission peak). However, a majority of the YSOs identified in N\,159 are found to be $<10\arcsec$ from the nearest clump's center, indicating their positions are significantly different than a random distribution and are correlated with the positions of the clumps.

As described above, the borders of the clumps are defined by a two-sigma detection threshold. Of the 36 clumps identified in our ATCA observations, 25 contain what have been identified as YSOs within their borders, and 5 of those have associated maser emission. Of the 25 clumps containing YSOs, 19 have at least 1 high-mass YSO ([8.0] $\leq$ 8), while the 6 others have only intermediate mass ([8.0] $>$ 8.0) YSOs. Only 1 of the candidate clumps contains a YSO.

In many cases, the YSO positions as determined from their IR emission is offset from the emission peak of the clump determined from the 0$^{\rm{th}}$ moment (integrated intensity) map of the HCO$^{+}$ emission but is still located within the contours of the clump. Recent modeling has shown that massive YSOs will form at the densest centers of their natal parsec-scale clouds \citep[e.g.,][]{krum10}, so the YSOs significantly offset from the peak of the clump may be more evolved than those more coincident with the core. The hypothesis that a YSO located near the center of a clump is evolutionarily younger than those located near the clump edge is observationally supported by our finding that most YSOs (4/5) with maser emission -- thought to be a signpost of early massive YSO evolution -- are located at or near the emission peaks of the surrounding clump.

If the offset between YSO position and clump center is evolutionary, then it might be expected for clump properties to change as a function of this offset. \citet{hill05} noted a number of Galactic massive star-bearing cores/clumps with star formation tracers that are offset from the peak of the cloud, but find no correlation between the clump's properties and the distance between the star formation tracer and the clump's emission peak. However, because the clumps they studied have sizes that range between $\sim0.1$ and $2$ pc, the distance between the clump center and the YSO may be influenced by the size of the clump as a whole. In other words, small YSO-bearing clumps will necessarily have YSOs close to their centers. To remove the dependence on clump size, we have determined a `distance factor' that takes into account both the projected YSO-clump center distance and the size of the clump. The distance factor, $\delta$, is defined as

\begin{equation}
\delta = \frac{C}{C+E},
\end{equation}

\noindent where $C$ is the projected distance between the YSO and the nearest clump emission peak, and $E$ is the projected distance between the YSO and the nearest clump edge. $E$ is defined as positive for YSOs within the borders of a clump and negative for YSO outside a clump. Under this definition, YSOs at the emission peak of a clump have distance factors of 0, those located directly on the edge have distance factors of 1, and YSOs outside of clumps have distance factors of $>1$. For clumps containing more than one emission peak, Table 3 reports the smallest value of $\delta$.

The HCO$^{+}$ clumps detected in this survey comprise four distinct classes (N, IM, CM, and EM) and are defined as follows:

\begin{enumerate}

\item \textbf{N (\underline{N}o YSO)}: clumps with no observed indications of recent intermediate or high-mass star formation.

\item \textbf{IM (\underline{I}ntermediate-\underline{m}ass YSO)}: clumps containing ($\delta \le 1.0$) at least one intermediate mass YSO and no high-mass YSOs.

\item \textbf{CM (\underline{C}enter \underline{m}assive YSO)}: clumps containing at least one massive YSO that is closer to the clump's center than its edge, i.e. $\delta \le 0.5$.

\item \textbf{EM (\underline{E}dge \underline{m}assive YSO)}: clumps with at least one massive YSO that is close to the clump's edge, i.e. $0.5 < \delta \le 1.0$ and no massive YSOs with $\delta \le 0.5$.

\end{enumerate}

In the following section, we compare the properties of the clumps within each of these classes.

\subsection{Relationship Between Clump Properties and Star Formation Activity}

For the purposes of comparing the properties of the clumps in each group, we restrict ourselves to the most robustly-determined properties, and thus exclude candidate clumps from the following analysis unless otherwise noted. Histograms comparing the M$_{\rm{LTE}}$ distributions of the different classes of clumps (Figure 8) show that the starless clumps are in general less massive than those associated with current star formation. The clumps without YSOs range in M$_{\rm{LTE}}$ between $2.6 \times 10^{2}$ and $2.3 \times 10^{3}$ M$_{\sun}$, with a mean mass of $1.5 \times 10^{3}$ M$_{\sun}$; YSO-bearing clumps have a mass range of $8.0 \times 10^{2}$ $-$ $3.3 \times 10^{4}$ M$_{\sun}$ and an average mass of $7.0 \times 10^{3}$ M$_{\sun}$. The difference in mass distributions is amplified when clumps of class IM are excluded -- clumps with high-mass YSOs (CM and EM) have an average mass of $7.9 \times 10^{3}$ M$_{\sun}$. We used the Student's t-test to statistically determine the significance of the differences in the mass distributions between the clump types; Table 4 reports the results. For each pair of clump types, we list the probability that the two populations do not have different means. Values below $\sim$0.05 (5$\%$) indicate significant differences in population distribution means. Note that the hypothesis that the clumps with and without YSOs come from the same population is rejected to a $>99\%$ confidence level.

The dearth of low-mass clumps (M $\lesssim 2 \times 10^{3}$ M$_{\sun}$) that are currently forming high-mass stars suggests that the mass of a clump is directly related to its massive star-forming capability. This is certainly true at the most basic level as a star could not form from clump material with a mass exceeding that of the natal clump. But we find that even the least massive clumps in this study have ample material to form a massive star, indicating there is an inherent property to the most massive clumps that makes them more favorable to forming massive stars. Recall that because we are only sensitive to the most massive YSO population members, the clumps we identify as being without YSO may not truly be free of star formation, and may be forming lower mass stars that our star formation tracers are insensitive to.

The finding of a lower mass for clumps without massive YSOs supports the findings of \citet{hill05}, who studied a large sample of Galactic, $0.1-2$ pc-sized `cores' (clumps under the terminology adopted here) in massive star formation regions. Cold, dense cores of dust were detected using 1.2-mm continuum observations, and a gas-to-dust ratio was assumed in order to estimate the cores' masses. They found that cores without maser or UC\ion{H}{2} emission (their primary tracers of star formation) have a median mass of $2.8 \times 10^{2}$ M$_{\sun}$, while cores with either masers or UC\ion{H}{2} regions have a median mass of $1.0 \times 10^{3}$. Moreover, within their sample, clumps without star formation are on average smaller in size than those with. Because many of our clumps are unresolved, the size distribution of the clumps in our sample is less secure. However, we also find that starless clumps tend to be smaller than those with YSOs, and that significantly more YSO-less clumps ($\sim80\%$) cannot be deconvolved from the beam compared to YSO-bearing clumps ($\sim20\%$).

We can estimate clump surface densities of our clump sample using M$_{\rm{LTE}}$ along with the clump surface area; YSO-less clumps have surface densities on the order of $0.1$ g cm$^{-2}$, while the clumps with massive YSOs have an average surface density of $\sim1$ g cm$^{-2}$. Differences in surface densities may reflect differences in volume densities -- YSO-bearing clumps (class CM, EM, and I) are denser than those without YSOs (class N). The higher volume density of YSO-bearing clumps is supported by the relatively high HCN/HCO$^{+}$ flux ratio seen in clumps with YSOs. For the reasons discussed in Section 3.2, HCN likely traces higher densities than HCO$^{+}$, and because clumps radially decrease in density, the two molecules likely trace different volumes of gas. Therefore a higher HCN/HCO$^{+}$ ratio is expected from structures containing high volume density gas. Clumps with the lowest densities may have little to no HCN emission at all. Indeed, while the HCN/HCO$^{+}$ ratio is low for group N clumps (0.19), the average HCN/HCO$^{+}$ flux ratio for all YSO-bearing clumps is 0.27, and is even higher for CM clumps (0.36).

In a recent study of the role of radiation feedback on the fragmentation and star formation within gas clouds, \citet{krum10} find that only clouds with a high surface density ($\Sigma > 1$ g cm$^{-2}$) are able to form massive stars. In a radiation-hydrodynamic simulation of a 100 M$_{\sun}$ cloud with an initial surface density of $\Sigma = 0.1$ g cm$^{-2}$ (typical of Galactic low-mass star forming regions), the cloud fragments strongly, forming a number of low-mass stars and no massive stars. Conversely, clouds with higher surface densities ($\Sigma > 1$ g cm$^{-2}$) fragment less, and put most their mass into massive stars, even with a total cloud mass of only 100 M$_{\sun}$.  The observations presented here are consistent with the conclusions of the \citet{krum10} simulation -- the ability of a clump to form a massive star is determined by its surface density (or by extension volume density).

In summary, we find that the clumps forming YSOs are physically larger, more massive, and denser than clumps that are not. How exactly the starless clumps fit into the picture of massive star formation is unclear. One possibility is that they are the evolutionary precursors to the clumps actively forming stars. They may represent the earliest stages of massive star formation, before the formation of a YSO massive enough to have been detected in any LMC YSO surveys. In this case, their lower masses, sizes, and density could be attributable to their youth, and they may grow in size, mass, and density if they accumulate more material from the molecular cloud in which they reside. Alternatively, it may be that these clumps are never destined to form massive stars at all. Distinguishing the two scenarios is difficult, and it may be that both hypotheses are true. For example, YSO-less clump 6 of N\,159 is more massive than several of the clumps currently harboring massive YSOs, suggesting it could eventually form a star. Conversely, given the dearth of low-mass clumps with YSOs, clumps with masses of only a few $10^{2}$ M$_{\sun}$ may never form massive stars at all and may only form low-mass stars. In either scenario, because clumps with masses $>2 \times 10^{3}$ M$_{\sun}$ almost always contain massive YSOs, the onset of massive star formation within these massive clumps occurs quickly; massive clumps do not have a long YSO-less stage.

\section{Clump Dissipation and Destruction}
\subsection{Evidence for Clump Dissipation}
Along with a correlation between a clump's properties and the presence or absence of a YSO, there appears to also be a correlation with the location of a YSO within the clump. Clumps with YSOs more centrally located generally have larger masses than clumps with YSOs closer to their edges (Figure 8). A similar correlation is also apparent in the HCN flux of the clumps (Figure 9); all CM class clumps have high HCN fluxes and higher HCN/HCO$^{+}$ ratios. On the other hand, little to no HCN is detected towards clumps of class N, IM, or EM. Recall that we find that nearly all the maser-associated YSOs, possibly the youngest in our sample, are located very close to their clumps' peaks. This then implies that a massive YSO is first formed at an emission peak, where the clump densities were high and a core collapsed to form a star. YSOs located at the edges of a clump could represent more advanced stages of evolution where the YSOs have either moved away from the dense cores they formed in or have dissipated parts of their immediate surroundings, leaving only the parts of the clump far enough away or with high enough densities to have not been destroyed.

Two alternative, non-evolutionary explanations are possible for the differences in mass distribution between clumps with YSOs at the centers and those with YSOs at the edges. First, higher mass clumps could be more centrally peaked while intermediate-mass clumps are more flocculent. YSOs would then form preferentially in the higher-mass clumps where the bulk of the gas is located and the densities are highest. In less massive clumps, the mass distribution may be less peaked, and YSOs could form in a more distributed fashion. This scenario also accounts for the low HCN flux in intermediate-mass clumps -- because it traces higher densities, HCN would only be detected towards the most centrally-peaked, densest clumps' centers. The second possibility is that because the most massive clumps also tend to be larger, a YSO has a greater probability of being located near the center of a more massive clump simply because the inner region of a massive clump has more surface area. This could result in more massive clumps being classified as CM than EM. While the largest clumps are of the CM class, the size distributions of CM and EM clumps do not differ significantly (see Figure 10). Our finding that masers are preferentially found at the centers of clumps may not necessarily argue against these scenarios. There may be a bias towards forming masers only in the most massive clumps or the densest regions of clump. Indeed, the existence of a maser requires high columns that can shield the masing molecular material from the star's ionizing radiation \citep[e.g.,][]{genz81}.

There are 20 YSOs within our ATCA-imaged area ($38\%$) that are not located within dense clumps. Or, said more precisely, are \textit{not currently} located within dense clumps, as they likely earlier formed within one. YSOs not located within clumps therefore could represent the more advanced stages of YSO evolution after a YSO has dissipated its surroundings. The dearth of clumps with masses $\lesssim 10^{3}$ M$_{\sun}$ harboring massive YSOs  suggests that after the onset of clump disruption, destruction of the clump proceeds rapidly; a clump that is undergoing the process of dispersal spends very little time in a low-mass phase. The result is essentially a bimodal distribution of massive YSOs -- those in massive (M $\gtrsim$ 10$^{3}$ M$_{\sun}$) clumps and those not in clumps at all.

We find that the YSOs located outside clumps tend to be visible at near-IR and even optical wavelengths, consistent with little circumstellar obscuration. Conversely, most sources in clumps are not visible at optical wavelengths, suggesting they are surrounded by a significant enough column to extinct the short wavelength emission. Indeed, we find that the \citet{chen09} Type classification scheme is very well correlated with the association of a YSO with dense molecular clumps, is a good indicator of a source's `embeddedness,' and can be considered evolutionary in origin. Figure 11 shows the proportion of YSOs within and outside of clumps that belong to each YSO Type from \citet{chen09, chen10}. Note that YSOs within clumps tend to be of Type I, I/II or II while YSOs outside of clumps are of Type II, II/III, or III. 

A YSO's propensity to be located within a clump is also well correlated with the presence of silicate absorption at 10 $\mu$m in its infrared spectrum. While the silicate spectral feature could potentially originate from any silicate-bearing dust particles located between the YSO and the observer, the bulk of this material is located in the YSO's immediate circumstellar environment. In time, the YSO dissipates its surroundings, removing circumstellar gas and dust, thus weakening the strength of the silicate absorption feature. There is little contribution to the silicate feature's depth from foreground Galactic dust. \citet{schw91} find that the extinction from Milky Way dust in the direction of the LMC is low; $A_{V}=$ 0.2--0.5, corresponding to a 10 $\mu$m silicate feauture optical depth $\tau_{10}$ of 0.01--0.03. The $\tau_{10}$ of the silicate absorption present in the YSOs' spectra measure between $\sim0.2$ and 2.0, meaning Galactic dust accounts for at most $15\%$ of the silicate feature's optical depth.

Many of the YSOs in this study were observed in SL09, who measured the strength of the silicate feature relative to the continuum using PCA. Table 3 reports the PCA-determined silicate strength for the SL09 YSOs, and Figure 12 shows the close correspondence between silicate strength (circumstellar dust) and HCO$^{+}$ flux (circumstellar molecular gas). Figure 13 shows the distribution of PCA silicate strengths for YSOs within ($\delta \le 1.0$) and outside ($\delta > 1$) of a clump. The silicate feature is strongest for sources that are located centrally within a clump; all YSOs with $\delta \le 0.5$ have PCA silicate strengths $> 0$. This is in contrast with the YSOs located outside of clumps: 4/9 of their members have silicate strengths $< 0$. Clearly the strength of the silicate feature is dependent on the presence of a circumstellar clump.

The correlation is highly suggestive that the silicate absorption feature and HCO$^{+}$ emission originate from the same medium; the silicate absorption may be originating from dust on the same scale as the clumps. Unlike interstellar ices, silicates do not require cold, dense, radiation shielded environments; they are detected even towards diffuse interstellar clouds \citep[e.g.,][]{bowe98}. The silicates contributing to the 10 $\mu$m feature may not be isolated to the YSO's immediate envelope or cold core, and may reside in the parsec-scale clump or even in the larger-scale, less dense molecular cloud.

\subsection{The Time Scale of Clump Disruption}

If the strength of the silicate feature is indeed an accurate indicator of the presence of a circumstellar clump around a YSO, given a complete sample of LMC YSO mid-IR spectra, we can estimate the number of YSOs located within clumps and therefore the relative time a YSO spends in clump-embedded and clump-free evolutionary stages. The SL09 IRS spectral catalog consists of nearly every YSO in the LMC that meets the criteria for being a candidate massive YSO in the GC09 catalog, and so we use it to statistically estimate the clump disruption time. In particular, because the average silicate strength of the entire SL09 massive YSO catalog is $\sim$0, lower than the average value for YSOs located within clumps ($\sim0.2$), it can be inferred that many YSOs from SL09 are not in clumps and that massive YSOs must spend a non-zero fraction of their lives not located within dense clumps. We estimate the number of YSOs in the SL09 sample located within clumps using the procedure outlined below.

We fit the distribution of silicate strengths for YSOs observed in clumps as a Gaussian with an average value of 0.23 and a FWHM of 0.5 (bottom panel of Figure 13). YSOs not in clumps have silicate strengths less than $\sim0.3$, and we therefore make the assumption that all YSOs in SL09 with silicate strengths greater than 0.3 are within clumps. Note that the SL09 silicate strength distribution (top panel of Figure 12) appears to have a larger right-hand wing than left-hand wing, suggesting that the entire distribution is composed of two separate distributions -- one with a higher average silicate strength than the other (i.e., YSOs within and outside of clumps, respectively). To this end, we fit the entire SL09 catalog of silicate strengths using a summation of two Gaussians: 1) the first (the right Gaussian curve in the top panel of Figure 13) is representative of the clump-embedded YSO population whose central position and width is defined by the aforementioned Gaussian fit to YSOs observed in clumps and whose height is determined by a fit to all SL09 YSOs with large ($>0.3$) PCA silicate strengths, and 2) the second is representative of YSOs not in clumps, whose height, central wavelength, and width are allowed to vary as free parameters. The implied distribution of YSOs not in clumps is the left Gaussian curve shown in the top panel of Figure 13, and has a central value of $-0.10$ and a FWHM of 0.33.

It is suggested in SL09 that sources with PCA silicate strengths of greater than 0.2 are embedded, while those with weaker PCA strengths are not. However, these results suggest that there are a number of YSOs embedded in clumps with silicate strengths less than 0.2. Therefore the total number of SL09 YSOs still embedded in their natal clumps is likely larger than is implied in SL09 ($20\%$). The Gaussian fits to the entire SL09 catalog presented here suggests that $34\%$ of all the massive YSOs in SL09 are within clumps; i.e., a massive YSO spends $\sim\frac{1}{3}$ of its life embedded in a clump during the YSO evolutionary stages probed by SL09.

GC09 (the catalog from which SL09 is based) identified YSOs via their \textit{Spitzer} photometry, meaning SL09 does not have the spectra for YSOs outside of the GC09 photometric selection criteria. Specifically, GC09 misses the youngest, most embedded YSOs (too dim at \textit{Spitzer} wavelengths) and the most evolved YSOs that have substantially dissipated their surroundings (too blue in color to meet GC09 color selection criteria). Figure 14, a graphical representation of the stages in massive star formation, shows the time period explored by SL09, $t_{\rm{SL09}}$. The highly embedded stage before \textit{Spitzer} detection is identified as $t_{\rm{A}}$, while the more evolved stage immediately following its identification with \textit{Spitzer} is $t_{\rm{D}}$. We divide $t_{\rm{SL09}}$ into clump-embedded and clump-free stages, $t_{\rm{B}}$ and $t_{\rm{C}}$, respectively; from the analysis we present above, we estimate that $t_{\rm{B}} =  \frac{1}{3} \times t_{\rm{SL09}}$ and $\frac{t_{\rm{B}}}{t_{\rm{C}}}\approx \frac{1}{2}$.

The total clump disruption time, the time between when a massive star forms in the clump and when the clump would no longer be detectable by our ATCA observations, is represented in the timeline as $t_{\rm{SF}}$ = $t_{\rm{A}}+t_{\rm{B}}$, where $t_{\rm{A}}$ is the time between the formation of a massive YSO and its detection by \textit{Spitzer}. Our determination of $t_{\rm{B}}$ sets a lower limit on $t_{\rm{SF}}$ such that $t_{\rm{SF}} \ge  \frac{1}{3} \times t_{\rm{SL09}}$ yrs. During $t_{\rm{SF}}$, the clump is able to form cluster members, but after the dense clump is dispersed, massive star formation -- possibly all star formation -- within the clump will cease.

Determining $t_{\rm{SF}}$ is difficult as there have been to date no published surveys of the entire LMC targeting the most embedded YSOs that are not detectable at near-  or mid-IR wavelengths, those in $t_{\rm{A}}$. Comparing this star formation stage to low-mass YSOs, these objects are comparable to the so-called `Class 0' sources that are strong blackbody emitters in the far-IR. To estimate $t_{\rm{A}}$, we use the findings of \citet{sewi10}, who conducted a search for Class 0 YSOs in a $2^{\circ}\times8^{\circ}$ strip of the LMC (approximately $25\%$ of the LMC's total surface area) with the \textit{Herschel Space Observatory} \citep{pilb10}. The survey identified $\sim80$ YSO candidates that were not identified in \textit{Spitzer} studies in this subregion of the galaxy. Extrapolating to include the entire LMC, we estimate there may be $\sim350$ YSOs that are too embedded to have been detected in the GC09 and W08 \textit{Spitzer} surveys but can be identified with \textit{Herschel}. The \citet{sewi10} survey is likely sensitive to a similar YSO stellar mass range as the GC09 and W08 $\textit{Spitzer}$ surveys ($M_{star} \gtrsim5$ M$_{\sun}$), so the two surveys are likely probing the same population of YSOs at two different stages of evolution. We therefore use the ratio of the number of YSOs detected by \textit{Spitzer} ($\sim1300$) to that predicted to be detected by only \textit{Herschel} ($\sim350$) to estimate $t_{\rm{A}} \approx \frac{1}{4} \times t_{\rm{SL09}}$ and $t_{\rm{SF}}$ = $t_{\rm{A}}+t_{\rm{B}} \approx \frac{7}{12} \times t_{\rm{SL09}}$.

To estimate a numeric value for $t_{\rm{SL09}}$, we determine the total mass of the LMC's \textit{Spitzer}-detected massive YSOs and assume a formation rate for that population. A Salpeter IMF \citep[$dN/dM = M^{-\alpha}$; $\alpha=2.35$;][]{salp55} predicts massive stars (M $\geq$ 8 M$_{\sun}$) account for $\sim10\%$ of the total mass of a stellar population. Therefore, the formation rate of massive stars, $MSFR$, is $10\%$ the star formation rate of all stars, $SFR$, i.e. $MSFR$ = 0.1 $\times$ $SFR$. The total mass of $N$ massive (8 M$_{\sun} < M <$ 50 M$_{\sun}$) stars is $M(\geq 8_{\sun})\approx16N$ M$_{\sun}$ in a Salpeter IMF population, so if one assumes a constant star formation rate, it then follows that it takes time 

\begin{equation}
t\ \rm{[yr]} = \frac{\textit{M}_{\textit{M}\geq8}}{\textit{MSFR}} = 160 \times \frac{\textit{N}}{\textit{SFR}\ \rm{[M_{\sun}\ yr^{-1}]}} 
\end{equation}

\noindent to form $N$ massive YSOs, where $SFR$ is the average total galactic star formation rate in M$_{\sun}$ yr$^{-1}$. Together, the GC09 and W08 studies have identified $N$=300 distinct massive YSO candidates, where massive is defined as above, [8.0]$\leq$ 8.0. We infer that the LMC's massive YSOs have a combined mass of $4.8\times10^{3}$ M$_{\sun}$. The LMC's global $SFR$ has been estimated previously by using the integrated H-$\alpha$, UV, and far-IR fluxes as star formation tracers. For example, \citet{calz07} find $SFR$s of 0.05 M$_{\sun}$ yr$^{-1}$ and 0.14 M$_{\sun}$ yr$^{-1}$ from the galaxy's total far-IR and H-$\alpha$ flux, respectively, while W08 reports a $SFR$s of 0.17 M$_{\sun}$ yr$^{-1}$ from the total UV flux. We adopt a typical literature $SFR$ of 0.1 M$_{\sun}$ yr$^{-1}$. Substituting these values of $N$ and $SFR$ into equation (4), we estimate $t_{\rm{SL09}}\approx5\times10^{5}$ yrs and $t_{\rm{SF}}\approx3\times10^{5}$ yrs. Note that because many of the massive YSOs identified in GC09 and W08 are likely multiple systems containing several massive stars, we may underestimate $M_{\textit{M}\geq8}$, and consequently $t_{\rm{SL09}}$ and $t_{\rm{SF}}$, by a factor of a few.

Massive stars disrupt their surroundings with ionizing and non-ionizing photons, stellar and pre-stellar winds, and eventually supernovae. \citet{tan01} simulated the destruction from ionizing radiation, stellar winds, and radiation pressure of a 0.1 pc, $10^{3}$ M$_{\sun}$ clump by a forming cluster. They find a destruction time of $0.3-2$ Myrs, depending on the level of internal substructure of the clump, similar to the values we estimate here observationally. The shortest modeled destruction time, $3\times10^{5}$ yr was found for a uniform, quiescent cloud containing 3 massive stars ($M_{star}>8M_{\sun}$). Longer clump lifetimes are possible if the clump itself is highly clumpy; if most of the clump's mass is in dense cores with velocities set by virial equilibrium, the ionized gas that is destroying the clump can be contained through what they call ``turbulent mass loading'' for 1.5--2 Myrs. The clump destruction time we estimate from the LMC sample is comparable to or may exceed the destruction time of a uniform clump if we underestimate $t_{\rm{SL09}}$ for the reasons outlines above, suggesting that turbulent mass loading may indeed be an important process to confining the ionized gas and prolonging the life of the clump.

Important to the theory of massive star formation is how this clump lifetime compares to the free-fall time of the clump, $t_{ff} = \sqrt{3\pi/(32G\rho)} = 1.38\times10^{6}(n_{H}/10^{3}cm^{-3})^{-1/2}$ yr. For $n_{H} = 10^{5}$ cm$^{-3}$, typical of Galactic star-forming clumps and approximately equal to the critical densities of the dense gas tracers HCO$^{+}$ and HCN, this yields $t_{ff} = 1.4\times10^{5}$ yr. Our approximation of the clump destruction time presented above is admittedly crude (assumes an LMC $SFR$, requires an IMF extrapolation, etc.), however it gives a clump lifetime after the formation of a massive star of at least 2 times the free-fall time. Without a source of support against collapse, the clump should collapse in a free-fall time, implying a source of support against collapse such as turbulence or magnetic fields is required to allow the clump to last the several $t_{ff}$ estimated here.

\subsection{Clump Star Formation Efficiency}

The star formation efficiency (SFE) of the star-forming clumps, defined as the fraction of total mass converted into stars over the lifetime of a cloud, can be estimated by dividing the total mass of YSOs by the sum of the YSO and clump masses. The summed M$_{\rm{LTE}}$ of the massive YSO-bearing clumps is $1.5\times10^{5}$ M$_{\sun}$. As described in Section 6.2.2, by counting the number of massive YSOs and assuming a Salpeter IMF, we can estimate the total stellar mass within the YSO-bearing clumps, $\sim4.5\times10^{3}$ M$_{\sun}$. Because most -- possibly all -- of these YSOs are unresolved clusters containing several massive stars, the total YSO mass is more realistically a factor of several higher. From these total clump and YSO masses we estimate the SFE of the clumps to be $>3\%$. The SFE of $>3\%$ is on the high end of SFE's estimated for entire GMCs, typically $\sim1\%$ \citep[e.g.,][]{duer82} and higher than the global LMC SFE of $1\%$ estimated by \citet{fuku99}. Our estimate is more crude than Galactic determinations as we are unable to individually count cluster members, but is markedly larger than that of LMC GMCs as a whole, an expected trend as SFE should increase as one probes higher densities. Our clump SFE estimates are more typical of those estimated for Galactic embedded clusters such as Serpens, Rho Oph, NGC 1333, and NGC 2071 ($\sim10\%$) \citep[see][and references therein]{lada03}.

\section{Summary and Conclusions}

We have conducted a survey of the dense molecular material in active star formation regions in the LMC using the ATCA. The observations, the first systematic interferometric mapping of dense gas in the LMC, were centered on the peaks of the regions' GMCs, where we identified a total of 46 dense molecular clumps with the \textsf{CPROPS IDL} package. Much of the molecular material is associated with signposts of on-going intermediate and massive star formation including maser emission, compact \ion{H}{2} regions, and bright IR point sources. We have categorized the clumps by the extent and nature of their star formation content, and find correlations between the categories and clump properties such as mass and size. Our primary conclusions are summarized below:

\begin{enumerate}

\item The gas within a GMC is clumpy with substructures that are directly revealed by the high volume density tracers HCO$^{+}$ and HCN. The ratio of the luminosity of the two transitions, $F_{\rm{HCN}}$/$F_{\rm{HCO^{+}}}$ decreases with increasing clump mass, column density, and star formation activity, with HCN emission generally only being detected from the most massive and densest clumps. The HCN/HCO$^{+}$ ratio is $\sim1/3$ in massive clumps (M $>\ 10^{3}$ M$_{\sun}$), and on average a factor of two lower in their less massive counterparts. The least massive clumps display no detectable HCN emission. Supported by the smaller physical scale from which HCN emission emanates, HCN likely is tracing higher volume densities than HCO$^{+}$, and its observation is restricted to the higher density centers of only the most dense clumps, i.e., those most capable of forming stars.

\item Determined from either the virial theorem or an integration over the surface area of the estimated column density, the clumps' masses span several orders of magnitude from a few $10^{2}$ M$_{\sun}$ to $\sim3 \times 10^{4}$ M$_{\sun}$. A fit to the integral of the mass function is consistent with a power law distribution with index $\alpha$ = $-2$. We determine the clumps to have radii between $\lesssim$1 pc and $\sim2$ pc and surface densities of 0.1 -- 1 g cm$^{-2}$.

\item The clumps are found to contain varying levels of current star formation, and we categorized them into four distinct classes: clumps with no on-going star formation (N), clumps containing only intermediate-mass YSOs (IM), clumps high high-mass YSOs located near their emission peaks (CM), and clumps with high-mass YSO located close to the clump edge (EM).

\item Clumps with and without signs of recent or current star formation differ in their physical properties. Massive YSO-bearing clumps tend to be larger ($\gtrsim$1 pc), more massive (M $\gtrsim10^{3}$ M$_{\sun}$), and have higher surface densities ($\sim1$ g cm$^{-2}$), while clumps without signs of star formation are smaller ($\lesssim$1 pc), less massive M $\lesssim$ M$10^{3}$M$_{\sun}$, and have lower surface densities ($\sim0.1$ g cm$^{-2}$). The implication is that the ability of a clump to form massive stars is determined by its physical attributes.

\item The dearth of massive (M $>$ 10$^{3}$ M $_{\sun}$) clumps without signs of massive star formation suggests the onset of star formation is rapid within a masive clump and occurs on a timescale significantly shorter than the clump's lifetime.

\item Clumps with centrally-located massive YSOs (CM) generally have masses 2--3 times that of clumps with massive YSOs only on their edges (EM). We suggest that the difference may be evolutionary, with YSOs being born at the centers of clumps but becoming displaced from the clumpÕs emission peak with time, while removing $1/2$ to $2/3$ of the clump mass by the time it is located at the clump edge. The lack of clumps harboring massive YSOs with masses $\lesssim10^{3}$ M$_{\sun}$ suggests that after the onset of clump disruption, destruction of the clump proceeds rapidly; a clump that is undergoing the process of dispersal spends very little time in a low-mass phase. The result is essentially a bimodal distribution of massive YSOs -- those in massive clumps and those not in clumps at all.

\item There is a strong correlation between the intensity of HCO$^{+}$ and HCN emission and the strength of the 10 $\mu$m silicate absorption feature seen in the mid-IR spectra of YSOs. Objects located within dense clumps have the strongest silicate features while those outside of clumps show little to no silicate absorption. The close correspondence indicates the absorbing particles responsible for the silicate feature may reside in the parsec-scale structures defined by the clumps.

\item Using the large sample of LMC massive YSO spectra presented in SL09, we estimate from the strength of the sample's silicate features that $34\%$ of the YSOs in the catalog are embedded in clumps. By adopting a global $SFR$ for the LMC of 0.1 M$_{\sun}$ yr$^{-1}$, we estimate a massive YSO spends at least $5\times10^{5}$ yrs in the evolutionary stages present in the \textit{Spitzer} YSO catalogs and derive a total clump destruction time after the onset of massive star formation of at least $3\times10^{5}$ yrs.

\end{enumerate}

\acknowledgments
The authors wish to thanks the anonymous referee for comments that greatly enhanced the paper. This work makes use of data collected by the \textit{Australia Telescope Compact Array}. The Australia Telescope is funded by the Commonwealth of Australia for operation as a National Facility managed by CSIRO. This work is based in part on observations made with the \textit{Spitzer Space Telescope}, which is operated by the Jet Propulsion Laboratory, California Institute of Technology under a contract with NASA. Support for this work was provided by NASA through an award issued by JPL/Caltech. J.P.S, L.W.L, and T.W. acknowledge support from NSF grant AST 08-07323. Part of the research described in this paper was carried out at the Jet Propulsion Laboratory, California Institute of Technology, under a contract with the National Aeronautics and Space Administration.

{\it Facilities:} \facility{ATCA, \textit{Spitzer} (IRS, IRAC, MIPS)}

\begin{deluxetable}{lcccccccc}
\tablewidth{0pt}
\rotate
\tablecolumns{8}
\tabletypesize{\tiny}
\tablecaption{Observations}
\tablehead{ 
\colhead{Region} & \colhead{RA of} & \colhead{Dec of} & \colhead{Observation Dates} & \colhead{Number of} & \colhead{HCO$^{+}$ beam size} & \colhead{Typical sensitivity} & \colhead{Spectral channel} \\
\colhead{} & \colhead{mosaic center} & \colhead{mosaic center} & \colhead{} & \colhead{Pointings} &  \colhead{} & \colhead{in 0.4 km s$^{-1}$ channel} & \colhead{width} \\
\colhead{} & \colhead{[h:m:s]} & \colhead{[d:m:s]} & \colhead{} & \colhead{} &  \colhead{[arcsec $\times$ arcsec]} & \colhead{[mJy beam$^{-1}$]} & \colhead{[km s$^{-1}$]} \\

}

\startdata
N\,105				& 05:09:51 & -68:53:35 & 2007 Oct		& 10	& $6.3\times7.1$	& 70	& 0.21	\\
N\,113				& 05:13:19 & -69:22:30 & 2007 Sept	& 9		& $5.8\times6.2$	& 110	& 0.21	\\
N\,159				& 05:39:51 & -69:45:35 & 2006 Sept	& 40	& $7.0\times6.0$	& 80	& 0.40	\\
N\,44 North (Region 1)	& 05:22:05 & -67:57:55 & 2008 Aug		& 36	& $5.5\times6.5$	& 80	& 0.42	\\
N\,44 South (Region 2)	& 05:22:55 & -68:04:14 & 2008 Aug		& 9		& $6.0\times7.2$	& 90	& 0.42	\\

\enddata
\end{deluxetable}

\clearpage

\begin{deluxetable}{ccccccccccl}
\tablewidth{0pt}
\rotate
\tablecolumns{10}
\tabletypesize{\tiny}
\tablecaption{Clump Properties}
\tablehead{ 
\colhead{Clump ID} & \colhead{RA (J2000)} & \colhead{Dec (J2000)} & \colhead{$\Delta$v} & \colhead{CPROPS Radius} & \colhead{M$_{\rm{vir}}$} & \colhead{M$_{\rm{LTE}}$\tablenotemark{a}} & \colhead{HCO$^{+}$ flux} & \colhead{HCN flux} & \colhead{Class\tablenotemark{b}} \\
\colhead{Number} & \colhead{[h:m:s]} & \colhead{[d:$\prime$:$\prime\prime$]} &  \colhead{[km sec$^{-1}$]} & \colhead{[pc]} & \colhead{[M$_{\sun}$]} & \colhead{[M$_{\sun}$]} & \colhead{[Jy km sec$^{-1}$]} & \colhead{[Jy km sec$^{-1}$]} & \colhead{}
}

\startdata
\cutinhead{\bf{N\,105}}
1	& 05:09:52.2	& --68:53:29	& 4.0 $\pm$ 0.2	& 1.7 $\pm$ 0.1	& 5.2 $\pm$ 0.5 $\times 10^{3}$	& 1.5 $\pm$ 0.2 $\times 10^{4}$	& 7.55 $\pm$ 0.41	& 3.43 $\pm$ 0.40	& CM, maser	\\ 
2	& 05:09:49.6	& --68:54:04	& 2.5 $\pm$ 0.4	&\nodata		& $\le1.0\pm0.1\times 10^{3}$	& 0.9 $\pm$ 1.9 $\times 10^{3}$	& 0.87 $\pm$ 0.06	& 0.29 $\pm$ 0.06	& CM		\\ 
3	& 05:09:52.2	& --68:53:02	& 2.9 $\pm$ 0.3	& 0.6 $\pm$ 0.4	& 9.4 $\pm$ 6.6 $\times 10^{2}$	& 2.3 $\pm$ 3.1 $\times 10^{3}$	& 1.86 $\pm$ 0.13	& 0.42 $\pm$ 0.13	& CM		\\ 
4	& 05:09:50.4	& --68:53:04	& 4.2 $\pm$ 0.1	& 1.5 $\pm$ 0.1	& 4.9 $\pm$ 0.5 $\times 10^{3}$	& 1.1 $\pm$ 0.2 $\times 10^{4}$	& 6.10 $\pm$ 0.28	& 2.14 $\pm$ 0.28	& CM		\\ 
5	& 05:09:51.5	& --68:52:47	& 1.7 $\pm$ 0.7	&\nodata		& $\le4.9\pm1.1\times 10^{2}$	& 0.6 $\pm$ 1.2 $\times 10^{3}$	& 0.53 $\pm$ 0.04	& 0.11 $\pm$ 0.04	& N			\\
\cutinhead{\bf{N\,113}}
1	& 05:13:25.1	& --69:22:46	& 4.7 $\pm$ 0.2	& 0.9 $\pm$ 0.2	& 3.8 $\pm$ 0.9 $\times 10^{3}$	& 1.2 $\pm$ 0.6 $\times 10^{4}$	& 7.11 $\pm$ 0.38	& 2.57 $\pm$ 0.39	& CM, maser	\\ 
2	& 05:13:21.2	& --69:22:42	& 5.3 $\pm$ 0.2	& 1.6 $\pm$ 0.1	& 8.6 $\pm$ 0.8 $\times 10^{3}$	& 9.4 $\pm$ 1.2 $\times 10^{3}$	& 5.88 $\pm$ 0.22	& 2.15 $\pm$ 0.22	& CM		\\ 
3	& 05:13:16.8	& --69:22:40	& 3.2 $\pm$ 0.2	& 1.0 $\pm$ 0.2	& 2.0 $\pm$ 0.4 $\times 10^{3}$	& 8.2 $\pm$ 3.3 $\times 10^{3}$	& 6.13 $\pm$ 0.37	& 1.94 $\pm$ 0.38	& I			\\ 
4	& 05:13:17.4	& --69:22:22	& 6.0 $\pm$ 0.1	& 0.5 $\pm$ 0.1	& 3.4 $\pm$ 0.7 $\times 10^{3}$	& 2.9 $\pm$ 1.2 $\times 10^{4}$	& 15.29 $\pm$ 0.70	& 8.17 $\pm$ 0.72	& CM, maser	\\ 
5	& 05:13:18.1	& --69:22:06	& 4.4 $\pm$ 0.4	& 0.7 $\pm$ 0.5	& 2.5 $\pm$ 1.9 $\times 10^{3}$	& 3.0 $\pm$ 4.3 $\times 10^{3}$	& 2.11 $\pm$ 0.10	& 0.64 $\pm$ 0.11	& EM		\\ 
6	& 05:13:18.5	& --69:21:50	& 2.5 $\pm$ 0.4	& \nodata		& $\le1.0\pm0.2\times 10^{3}$	& 1.7 $\pm$ 3.5 $\times 10^{3}$	& 1.37 $\pm$ 0.10	& 0.14 $\pm$ 0.11	& N			\\ 
\cutinhead{\bf{N\,159}}
1	& 05:39:37.4	& --69:46:09	& 3.9 $\pm$ 0.3	& 1.5 $\pm$ 0.2	& 4.2 $\pm$ 0.9 $\times 10^{3}$	& 3.8 $\pm$ 1.1 $\times 10^{3}$	& 2.26 $\pm$ 0.14	& 0.47 $\pm$ 0.14	& EM		\\ 
2\tablenotemark{c}	& 05:39:39.2	& --69:46:25	& 1.5 $\pm$ 1.1	& \nodata		& $\le3.6\pm0.8\times 10^{2}$	& 2.7 $\pm$ 5.7 $\times 10^{2}$	& 0.24 $\pm$ 0.04	& 0.10 $\pm$ 0.04	& N			\\ 
3	& 05:39:29.7	& --69:47:21	& 3.4 $\pm$ 0.4	& 0.7 $\pm$ 0.3	& 1.5 $\pm$ 0.7 $\times 10^{3}$	& 4.2 $\pm$ 3.6 $\times 10^{3}$	& 2.99 $\pm$ 0.26	& 0.96 $\pm$ 0.25	& EM, maser	\\ 
4	& 05:40:08.1	& --69:44:40	& 2.6 $\pm$ 0.6	& 0.2 $\pm$ 0.3	& 2.5 $\pm$ 3.9 $\times 10^{2}$	& 1.9 $\pm$ 5.7 $\times 10^{3}$	& 1.62 $\pm$ 0.12	& 0.39 $\pm$ 0.11	& EM		\\ 
5	& 05:40:04.4	& --69:44:34	& 3.2 $\pm$ 0.3	& 0.4 $\pm$ 0.2	& 7.5 $\pm$ 3.9 $\times 10^{2}$	& 7.1 $\pm$ 0.4 $\times 10^{3}$	& 4.23 $\pm$ 0.36	& 1.73 $\pm$ 0.35	& EM		\\ 
6	& 05:40:04.8	& --69:44:25	& 2.5 $\pm$ 0.5	& 0.8 $\pm$ 0.4	& 9.3 $\pm$ 6.2 $\times 10^{2}$	& 2.4 $\pm$ 2.4 $\times 10^{3}$	& 1.63 $\pm$ 0.18	& 0.60 $\pm$ 0.17	& N			\\ 
7	& 05:39:36.7	& --69:45:36	& 7.4 $\pm$ 0.2	& 2.5 $\pm$ 0.1	& 2.6 $\pm$ 0.2 $\times 10^{4}$	& 3.3 $\pm$ 0.3 $\times 10^{4}$	& 16.54 $\pm$ 0.59	& 5.80 $\pm$ 0.57	& CM, maser	\\ 
8	& 05:40:02.3	& --69:45:00	& 3.4 $\pm$ 0.6	& \nodata		& $\le2.0\pm0.4\times 10^{3}$	& 1.2 $\pm$ 2.4 $\times 10^{3}$	& 1.02 $\pm$ 0.08	& 0.34 $\pm$ 0.08	& N			\\ 
9\tablenotemark{c} & 05:39:41.5	& --69:46:56	& 2.3 $\pm$ 1.1	& \nodata		& $\le9.0\pm2.0\times 10^{2}$	& 0.7 $\pm$ 1.4 $\times 10^{3}$	& 0.61 $\pm$ 0.05	& 0.15 $\pm$ 0.05	& N			\\ 
10	& 05:39:40.7	& --69:46:33	& 3.2 $\pm$ 0.7	& \nodata		& $\le1.7\pm0.4\times 10^{3}$	& 0.8 $\pm$ 1.6 $\times 10^{3}$	& 0.72 $\pm$ 0.05	& 0.13 $\pm$ 0.05	& I			\\ 
11	& 05:39:58.6	& --69:45:03	& 3.2 $\pm$ 1.2	& \nodata		& $\le1.7\pm0.4\times 10^{3}$	& 1.2 $\pm$ 2.4 $\times 10^{3}$	& 0.65 $\pm$ 0.06	& 0.07 $\pm$ 0.06	& N			\\ 
12	& 05:39:41.7	& --69:46:11	& 3.9 $\pm$ 0.3	& 0.5 $\pm$ 0.2	& 1.4 $\pm$ 0.6 $\times 10^{3}$	& 4.4 $\pm$ 3.6 $\times 10^{3}$	& 3.65 $\pm$ 0.27	& 1.84 $\pm$ 0.26	& CM		\\ 
13\tablenotemark{c} & 05:39:31.3	& --69:45:24	& 1.7 $\pm$ 1.3	& \nodata		& $\le4.9\pm1.1\times 10^{2}$	& 4.2 $\pm$ 9.1 $\times 10^{2}$	& 0.45 $\pm$ 0.04	& 0.04 $\pm$ 0.04	& N			\\ 
14\tablenotemark{c} & 05:40:05.2	& --69:44:48	& 1.9 $\pm$ 1.7	& \nodata		& $\le6.0\pm1.3\times 10^{2}$	& 2.8 $\pm$ 6.1 $\times 10^{2}$	& 0.31 $\pm$ 0.03	& 0.08 $\pm$ 0.03	& N			\\ 
15\tablenotemark{c} & 05:39:37.0	& --69:46:15	& 2.7 $\pm$ 1.1	& 0.4 $\pm$ 1.3	& 5.5 $\pm$ 18.4 $\times 10^{2}$	& 0.8 $\pm$ 5.1 $\times 10^{3}$	& 0.58 $\pm$ 0.05	& 0.17 $\pm$ 0.05	& N			\\ 
16\tablenotemark{c} & 05:39:34.4	& --69:45:10	& 1.1 $\pm$ 1.4	& \nodata		& $\le2.2\pm0.5\times 10^{2}$	& 3.0 $\pm$ 7.1 $\times 10^{2}$	& 0.42 $\pm$ 0.04	& 0.09 $\pm$ 0.04	& N			\\ 
17	& 05:39:48.0	& --69:45:16	& 1.6 $\pm$ 0.9	& \nodata		& $\le4.1\pm0.9\times 10^{2}$	& 0.7 $\pm$ 1.4 $\times 10^{3}$	& 0.83 $\pm$ 0.06	& 0.16 $\pm$ 0.06	& N			\\ 
18	& 05:39:40.3	& --69:45:40	& 1.8 $\pm$ 1.5	& \nodata		& $\le5.3\pm1.2\times 10^{2}$	& 0.8 $\pm$ 1.7 $\times 10^{3}$	& 0.86 $\pm$ 0.05	& 0.20 $\pm$ 0.05	& N			\\ 
19	& 05:39:45.0	& --69:45:10	& 2.1 $\pm$ 0.4	& \nodata		& $\le7.6\pm1.7\times 10^{2}$	& 1.9 $\pm$ 3.9 $\times 10^{3}$	& 1.56 $\pm$ 0.11	& 0.43 $\pm$ 0.11	& N			\\ 
20	& 05:39:42.8	& --69:45:06	& 2.0 $\pm$ 0.8	&1.2 $\pm$ 0.4	& 9.2 $\pm$ 7.5 $\times 10^{2}$	& 1.2 $\pm$ 0.9 $\times 10^{3}$	& 0.88 $\pm$ 0.08	& 0.10 $\pm$ 0.08	& N			\\ 
\cutinhead{\bf{N\,44 Region 1}}
1	& 05:22:03.4	& --67:57:44	& 4.5 $\pm$ 0.3	& 1.4 $\pm$ 0.2	& 5.2 $\pm$ 1.1 $\times 10^{3}$	& 4.5 $\pm$ 1.3 $\times 10^{3}$	& 3.10 $\pm$ 0.15	& 0.69 $\pm$ 0.18	& I			\\ 
2	& 05:22:05.7	& --67:57:51	& 4.9 $\pm$ 0.4	& 1.5 $\pm$ 0.2	& 6.7 $\pm$ 1.4 $\times 10^{3}$	& 3.7 $\pm$ 1.0 $\times 10^{3}$	& 2.28 $\pm$ 0.12	& 0.33 $\pm$ 0.14	& I			\\ 
3	& 05:22:02.1	& --67:57:50	& 3.3 $\pm$ 0.3	& 1.4 $\pm$ 0.2	& 2.8 $\pm$ 0.6 $\times 10^{3}$	& 3.5 $\pm$ 1.1 $\times 10^{3}$	& 2.98 $\pm$ 0.18	& 0.49 $\pm$ 0.21	& CM		\\ 
4	& 05:22:12.8	& --67:58:32	& 5.2 $\pm$ 0.4	& 1.0 $\pm$ 0.2	& 5.0 $\pm$ 1.3 $\times 10^{3}$	& 3.5 $\pm$ 1.4 $\times 10^{3}$	& 2.46 $\pm$ 0.14	& 0.85 $\pm$ 0.16	& CM		\\ 
5	& 05:22:08.5	& --67:58:06	& 2.3 $\pm$ 0.6	& \nodata		& $\le8.8\pm2.0\times 10^{2}$	& 0.9 $\pm$ 1.8 $\times 10^{3}$	& 0.68 $\pm$ 0.07	& 0.01 $\pm$ 0.08	& I			\\ 
6	& 05:22:12.3	& --67:58:14	& 4.3 $\pm$ 0.4	& 1.6 $\pm$ 0.2	& 5.6 $\pm$ 1.2 $\times 10^{3}$	& 5.5 $\pm$ 1.5 $\times 10^{3}$	& 3.03 $\pm$ 0.17	& 0.41 $\pm$ 0.20	& I			\\ 
7	& 05:22:07.8	& --67:58:27	& 4.2 $\pm$ 0.4	& 1.4 $\pm$ 0.2	& 4.7 $\pm$ 1.1 $\times 10^{3}$	& 4.7 $\pm$ 1.4 $\times 10^{3}$	& 2.65 $\pm$ 0.14	& 0.36 $\pm$ 0.16	& EM		\\ 
8\tablenotemark{c} & 05:22:00.4	& --67:57:38	& 1.1 $\pm$ 1.2	& \nodata		& $\le2.0\pm0.4\times 10^{2}$	& 2.5 $\pm$ 5.8 $\times 10^{2}$	& 0.44 $\pm$ 0.04	& 0.01 $\pm$ 0.05	& N			\\ 
9\tablenotemark{c} & 05:22:04.3	& --67:58:18	& 2.6 $\pm$ 0.9	& \nodata		& $\le1.1\pm0.2\times 10^{3}$	& 0.8 $\pm$ 1.5 $\times 10^{3}$	& 0.54 $\pm$ 0.05	& 0.03 $\pm$ 0.06	& N			\\ 
\cutinhead{\bf{N\,44 Region 2}}
1	& 05:22:54.3	& -68:04:29	& 3.9 $\pm$ 1.2	& \nodata		& $\le2.6\pm0.6\times 10^{3}$	& 1.2 $\pm$ 2.5 $\times 10^{3}$	& 0.76 $\pm$ 0.06	& 0.04 $\pm$ 0.09	& N			\\ 
2	& 05:22:51.8	& -68:04:16	& 1.7 $\pm$ 1.0	& \nodata		& $\le5.0\pm1.1\times 10^{2}$	& 2.6 $\pm$ 5.5 $\times 10^{2}$	& 0.45 $\pm$ 0.05	& 0.06 $\pm$ 0.07	& N			\\ 
3	& 05:22:56.8	& -68:04:10	& 2.3 $\pm$ 0.6	& \nodata		& $\le8.8\pm2.0\times 10^{2}$	& 0.9 $\pm$ 1.9 $\times 10^{3}$	& 0.85 $\pm$ 0.10	& 0.07 $\pm$ 0.16	& EM		\\ 
4	& 05:22:55.2	& -68:04:07	& 3.2 $\pm$ 0.4	& \nodata		& $\le1.7\pm0.4\times 10^{3}$	& 1.6 $\pm$ 3.3 $\times 10^{3}$	& 1.39 $\pm$ 0.12	& 0.04 $\pm$ 0.19	& EM		\\ 
5\tablenotemark{c} & 05:22:58.3	& -68:03:48	& 2.0 $\pm$ 1.2	& \nodata		& $\le7.1\pm1.6\times 10^{2}$	& 4.7 $\pm$ 9.9 $\times 10^{2}$	& 0.55 $\pm$ 0.06	& 0.13 $\pm$ 0.10	& I			\\ 
6\tablenotemark{c} & 05:22:53.9	& -68:04:32	& 1.2 $\pm$ 1.5	& \nodata		& $\le2.2\pm0.5\times 10^{2}$	& 2.2 $\pm$ 5.2 $\times 10^{2}$	& 0.24 $\pm$ 0.04	& 0.01 $\pm$ 0.06	& N			\\ 
\enddata

\tablenotetext{a}{Note that the M$_{\rm{LTE}}$ uncertainties quoted are statistical and do not include uncertainties in the absolute HCO$^{+}$ flux due to flux calibration ($20\%$) or HCO$^{+}$ abundance.}
\tablenotetext{b}{Clump classes: N -- No YSO in clump; CM -- Centralized massive YSO in clump; EM -- Massive YSO only on the edge of the clump; IM -- Intermediate mass YSO in clump, no massive YSO in clump. Clumps with previously-identified maser activity are also indicated.}
\tablenotetext{c}{Candidate clumps (See Section 3.1).}

\end{deluxetable}

\clearpage

\begin{deluxetable}{lccccccccl}
\tablewidth{0pt}
\rotate
\tablecolumns{10}
\tabletypesize{\tiny}
\tablecaption{YSO Properties}
\tablehead{ 
\colhead{YSO ID} & \colhead{RA (J2000)} & \colhead{Dec (J2000)} & \colhead{Clump YSO is} & \colhead{[8.0]} & \colhead{IRS} & \colhead{PCA silicate} & \colhead{Distance} & \colhead{HCO$^{+}$ Flux Density} & \colhead{Notes}\\
\colhead{} & \colhead{[h:m:s]} & \colhead{[d:$\prime$:$\prime$$\prime$]} &  \colhead{located in} & \colhead{} & \colhead{Group\tablenotemark{a}} & \colhead{strength} & \colhead{factor $\delta$} & \colhead{[Jy beam$^{-1}$ km s$^{-1}$)]} & \colhead{}
}

\startdata
\cutinhead{\bf{N\,105}}
050949.80--685402.1	& 05:09:49.80	& --68:54:02.1	& 2			& 7.80 $\pm$ 0.08	& \nodata	& \nodata	& 0.15	& 1.38 $\pm$ 0.15	& 		\\
050950.11--685349.4	& 05:09:50.11	& --68:53:49.4	& \nodata	& 9.35 $\pm$ 0.11	& \nodata	& \nodata	& 2.87	& --0.03 $\pm$ 0.15	& 		\\
050950.12--685426.9	& 05:09:50.12	& --68:54:26.9	& \nodata	& 8.16 $\pm$ 0.09	& \nodata	& \nodata	& 3.22	& 0.06 $\pm$ 0.15	& 		\\
050950.53--685305.5	& 05:09:50.53	& --68:53:05.5	& 4			& 5.24 $\pm$ 0.05	& P			& 0.18		& 0.21	& 3.00 $\pm$ 0.15	& 		\\
050952.26--685327.3	& 05:09:52.26	& --68:53:27.3	& 1			& 6.82 $\pm$ 0.07	& PE		& 0.57		& 0.31	& 3.81 $\pm$ 0.15	& maser	\\
050952.73--685300.7	& 05:09:52.73	& --68:53:00.7	& 3			& 5.86 $\pm$ 0.08	& PE		& 0.06		& 0.50	& 1.40 $\pm$ 0.15	& 		\\
050953.89--685336.7	& 05:09:53.89	& --68:53:36.7	& 1			& 7.85 $\pm$ 0.07	& P			& 0.40		& 0.68	& 0.88 $\pm$ 0.15	& 		\\
\cutinhead{\bf{N\,113}}
051315.73--692135.9	& 05:13:15.73	& --69:21:35.9	& \nodata	& 6.97 $\pm$ 0.06	& PE		& 0.06		& 3.20	& 0.58 $\pm$ 0.22	& 		\\
051317.30--692236.7	& 05:13:17.30	& --69:22:36.7	& 3			& \nodata	 		& \nodata	& \nodata	& 0.49	& 2.88 $\pm$ 0.22	& 		\\
051317.54--692208.5	& 05:13:17.54	& --69:22:08.5	& 6			& 9.45 $\pm$ 0.13	& \nodata	& \nodata	& 0.71	& 0.51 $\pm$ 0.22	& 		\\
051317.69--692225.0	& 05:13:17.69	& --69:22:25.0	& 4			& 5.62 $\pm$ 0.06	& PE		& 0.23		& 0.40	& 5.69 $\pm$ 0.22	& maser	\\
051318.26--692135.5	& 05:13:18.26	& --69:21:35.5	& \nodata	& 7.98 $\pm$ 0.07	& PE		& --0.25		& 2.04	& 0.60 $\pm$ 0.22	& 		\\
051319.14--692151.0	& 05:13:19.14	& --69:21:51.0	& 5			& 6.78 $\pm$ 0.07	& PE		& --0.03		& 0.61	& 1.43 $\pm$ 0.22	& 		\\
051320.75--692151.4	& 05:13:20.75	& --69:21:51.4	& \nodata	& 8.75 $\pm$ 0.10	& \nodata	& \nodata	& 1.83	& --0.02 $\pm$ 0.22	& 		\\
051321.43--692241.5	& 05:13:21.43	& --69:22:41.5	& 2			& 5.67 $\pm$ 0.07	& PE		& 0.15		& 0.42	& 2.79 $\pm$ 0.22	& 		\\
051325.09--692245.1	& 05:13:25.09	& --69:22:45.1	& 1			& 5.53 $\pm$ 0.06	& P			& 0.32		& 0.14	& 5.60 $\pm$ 0.22	& maser	\\
\cutinhead{\bf{N\,159}}
053929.21--694719.0	& 05:39:29.21	& --69:47:19.0	& 3			& 7.20 $\pm$ 0.07	& P			& 0.36		& 0.68	& 1.63 $\pm$ 0.25	& Type I; maser 	\\
053935.99--694604.1	& 05:39:35.99	& --69:46:04.1	& \nodata	& 6.84 $\pm$ 0.06	& PE		& --0.08		& 1.05	& 0.08 $\pm$ 0.25	& Type I 		\\
053937.04--694536.7	& 05:39:37.04	& --69:45:36.7	& 7			& 6.60 $\pm$ 0.07	& PE		& 0.68		& 0.37	& 3.57 $\pm$ 0.25	& Type I 		\\
053937.53--694609.8	& 05:39:37.53	& --69:46:09.8	& 1			& 5.82 $\pm$ 0.06	& PE		& 0.04		& 0.32	& 1.95 $\pm$ 0.25	& Type I/II		\\
053937.56--694525.3	& 05:39:37.56	& --69:45:25.3	& 7			& 6.13 $\pm$ 0.08	& PE		& 0.37		& 0.36	& 3.94 $\pm$ 0.25	& Type I 		\\
053940.78--694632.1	& 05:39:40.78	& --69:46:32.1	& 10		& 8.02 $\pm$ 0.07	& \nodata	& \nodata	& 0.64	& 0.79 $\pm$ 0.25	& Type II 		\\
053941.89--694612.0	& 05:39:41.89	& --69:46:12.0	& 12		& 5.93 $\pm$ 0.06	& PE		& 0.38		& 0.25	& 2.98 $\pm$ 0.25	& Type I/II 		\\
053943.74--694540.3	& 05:39:43.74	& --69:45:40.3	& 18		& 9.91 $\pm$ 0.05	& \nodata	& \nodata	& 0.96	& 0.07 $\pm$ 0.25	& Type III 		\\
053945.18--694450.4	& 05:39:45.18	& --69:44:50.4	& \nodata	& 7.44 $\pm$ 0.06	& PE		& 0.05		& 3.44	& --0.31 $\pm$ 0.25	& Type III 		\\
053947.68--694526.1	& 05:39:47.68	& --69:45:26.1	& \nodata	& 8.14 $\pm$ 0.03	& \nodata	& \nodata	& 2.37	& 0.07 $\pm$ 0.25	& Type II 		\\
053951.60--694510.5	& 05:39:51.60	& --69:45:10.5	& \nodata	& \nodata			& PE		& \nodata	& 4.92	& 0.48 $\pm$ 0.25	&  				\\
053952.60--694517.0	& 05:39:52.60	& --69:45:17.0	& \nodata	& 8.14 $\pm$ 0.03	& \nodata	& \nodata	& 2.31	& --0.06 $\pm$ 0.25	& Type III 		\\
053959.34--694526.3	& 05:39:59.34	& --69:45:26.3	& \nodata	& 6.54 $\pm$ 0.06	& PE		& --0.37		& 3.90	& --0.04 $\pm$ 0.25	& Type II 		\\
054004.39--694437.6	& 05:40:04.39	& --69:44:37.6	& 5			& 5.40 $\pm$ 0.06	& PE		& 0.04		& 0.63	& 2.64 $\pm$ 0.25	& Type III		\\
054009.49--694453.5	& 05:40:09.49	& --69:44:53.5	& \nodata	& 7.31 $\pm$ 0.07	& PE		& 0.29		& 1.55	& 0.38 $\pm$ 0.25	& Type I/II		\\
\cutinhead{\bf{N\,44 Region 1}}
052156.97--675700.1	& 05:21:56.97	& --67:57:00.1	& \nodata	& 10.03 $\pm$ 0.10	& PE		& 0.06		& 8.13	& --0.40 $\pm$ 0.37	& Type III 			\\
052159.6--675721.7	& 05:21:59.6		& --67:57:21.7	& \nodata	& 10.09 $\pm$ 0.24	& \nodata	& \nodata	& 3.25	& --0.06 $\pm$ 0.37	& Type III 			\\
052201.9--675732.5	& 05::22:01.9		& --67:57:32.5	& \nodata	& 9.07 $\pm$ 0.25	& \nodata	& \nodata	& 2.28	& --0.02 $\pm$ 0.37	&					\\
052202.0--675758.2	& 05:22:02.0		& --67:57:58.2	& \nodata	& 7.77 $\pm$ 0.03	& \nodata	& \nodata	& 1.29	& --0.13 $\pm$ 0.37	& Type II 			\\
052202.11--675753.6	& 05:22:02.11	& --67:57:53.6	& 3			& 7.68 $\pm$ 0.70	& PE		& 0.29		& 0.48	& 0.43 $\pm$ 0.37	& Type II 			\\
052203.30--675747.0	& 05:22:03.30	& --67:57:47.0	& 1			& 8.36 $\pm$ 0.11	& PE		& 0.15		& 0.49	& 1.26 $\pm$ 0.37	& Type I/II 			\\
052203.9--675743.7	& 05:22:03.9		& --67:57:43.7	& 1			& 8.98 $\pm$ 0.10	& \nodata	& \nodata	& 0.49	& 1.24 $\pm$ 0.37	& Type II 			\\
052204.8--675744.6	& 05:22:04.8		& --67:57:44.6	& \nodata	& 8.76 $\pm$ 0.20	& \nodata	& \nodata	& 1.15	& --0.13 $\pm$ 0.37	& Type II/III 			\\
052205.2--675741.6	& 05:22:05.2		& --67:57:41.6	& \nodata	& 8.61 $\pm$ 0.23	& \nodata	& \nodata	& 1.62	& --0.68 $\pm$ 0.37	&  					\\
052205.3--675748.5	& 05:22:05.3		& --67:57:48.5	& 2			& 8.13 $\pm$ 0.04	& \nodata	& \nodata	& 0.58	& 1.12 $\pm$ 0.37	& Type II 			\\
052206.28--675659.1	& 05:22:06.28	& --67:56:59.1	& \nodata	& 8.40 $\pm$ 0.06	& \nodata	& \nodata	& 6.66	& --0.19 $\pm$ 0.37	& Type II 			\\
052207.27--675819.7	& 05:22:07.27	& --67:58:19.7	& \nodata	& 8.00 $\pm$ 0.06	& PE		& --0.13		& 1.30	& 0.18 $\pm$ 0.37	& Type III 			\\
052207.32--675826.8	& 05:22:07.32	& --67:58:26.8	& 7			& 8.30 $\pm$ 0.08	& \nodata	& \nodata	& 0.45	& 0.84 $\pm$ 0.37	& Type II 			\\
052208.6--675805.5	& 05:22:08.6		& --67:58:05.5	& 5			& 9.01 $\pm$ 0.09	& \nodata	& \nodata	& 0.15	& 0.80 $\pm$ 0.37	&  					\\
052211.86--675818.1	& 05:22:11.86	& --67:58:18.1	& 6			& 8.95 $\pm$ 0.08	& \nodata	& \nodata	& 0.67	& 0.04 $\pm$ 0.37	& Type I 			\\
052212.24--675813.2	& 05:22:12.24	& --67:58:13.2	& 6			& 8.97 $\pm$ 0.10	& \nodata	& \nodata	& 0.12	& 1.55 $\pm$ 0.37	& Type II/III 			\\
052212.57--675832.3	& 05:22:12.57	& --67:58:32.4	& 4			& 5.08 $\pm$ 0.05	& SE		& 0.44		& 0.16	& 3.38 $\pm$ 0.37	& Type I 			\\
\cutinhead{\bf{N\,44 Region 2}}
052251.62--680436.6	& 05:22:51.62	& --68:04:36.6	& \nodata	& 8.60 $\pm$ 0.09	& \nodata	& \nodata	& 2.10	& 0.09 $\pm$ 0.22	& 						\\
052255.12--680409.4	& 05:22:55.12	& --68:04:09.4	& 4			& 7.64 $\pm$ 0.07	& PE		& --0.03		& 0.48	& 0.54 $\pm$ 0.22	& Type II 				\\
052256.79--680406.8	& 05:22:56.79	& --68:04:06.8	& 3			& 7.77 $\pm$ 0.07	& PE		& --0.16		& 0.68	& 0.79 $\pm$ 0.22	& Type II 				\\
052257.55--680414.1	& 05:22:57.55	& --68:04:14.1	& 3			& 8.60 $\pm$ 0.07	& \nodata	& \nodata	& 0.97	& 0.40 $\pm$ 0.22	& Type I/II 				\\
052259.0--680346.3	& 05:22:59.0		& --68:03:46.3	& 5			& 10.46 $\pm$ 0.14	& \nodata	& \nodata	& 0.95	& 0.60 $\pm$ 0.22	& Type II 				\\
\enddata

\tablenotetext{a}{IRS Spectral Classes: S -- Silicate absorption only; SE -- Silicate absorption and fine-structure lines; P -- PAH emission; PE -- PAH emission and fine-structure lines}
\tablenotetext{b}{When indicated, massive YSO Types for N\,159 and N\,44 taken from \citet{chen09} and \citet{chen10}, respectively.}
\end{deluxetable}

\clearpage

\begin{deluxetable}{lcccccc}
\tablewidth{0pt}
\rotate
\tablecolumns{6}
\tabletypesize{\tiny}
\tablecaption{Student's t-test Analysis of Clump M$_{\rm{LTE}}$ Distributions}
\tablehead{ 
\colhead{} & \colhead{CM} & \colhead{EM} & \colhead{IM} & \colhead{CM+EM} & \colhead{CM+EM+IM}
}

\startdata
N		& 0.01				& 0.004				& 0.01		& 0.03				& 0.03\\
CM		& \nodata			& 0.06				& 0.08		& 0.37				& 0.17\\
EM		& \nodata			& \nodata			& 0.98		& 0.18				& 0.26\\
IM		& \nodata			& \nodata			& \nodata	& 0.21				& 0.30\\
\enddata
\tablecomments{The probability that the two clump populations (given by the row and column) do not have different means. Candidate clumps are not included in this calcualtion. Clump classes are defined in Section 5.}
\end{deluxetable}

\clearpage

\begin{figure}[t]
\begin{center}
\includegraphics[width=0.75\textwidth]{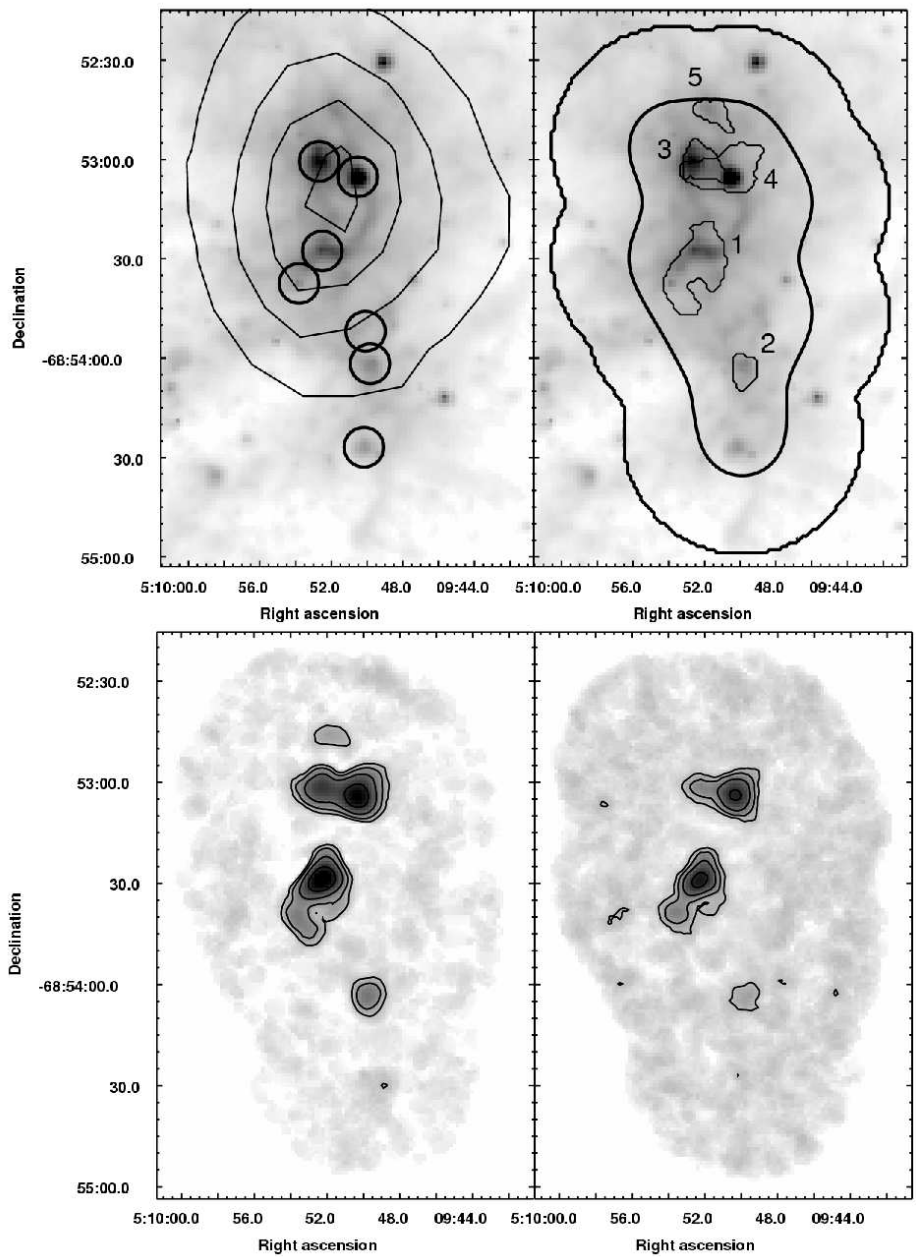}
\caption{\textit{Spitzer} 5.8 $\mu$m, MAGMA CO and ATCA HCO$^{+}$ and HCN images of N\,105. (Top left) Gray scale 5.8 $\mu$m image overlaid with MAGMA CO contours with the positions of YSOs marked with circles. Coutours are at 4$\times n$ (n=1,2,3,4). (Top right) Gray scale 5.8 $\mu$m image with the $50\%$ sensitivity and observational boundary of the ATCA HCO$^{+}$ mosaic in thick contours; clump boundaries are indicated with thin contours. (Bottom left) The 0$^{\rm{th}}$ moment (integrated intensity) of the ATCA HCO$^{+}$ masked version of the data cube (to reduce noise) in both gray scale and coutour. Contour levels are at $2\sigma\times2^{n}$ ($n$ = 0, 1, 2, 3, 4...). Note that the lowest contour represents a 2-sigma detection and that each successive contour represents a factor of two increase in signal-to-noise. (Bottom right) Same as the bottom left panel, but for HCN.}
\label{f1}
\end{center}
\end{figure}

\clearpage

\begin{figure}[t]
\begin{center}
\includegraphics[width=0.95\textwidth]{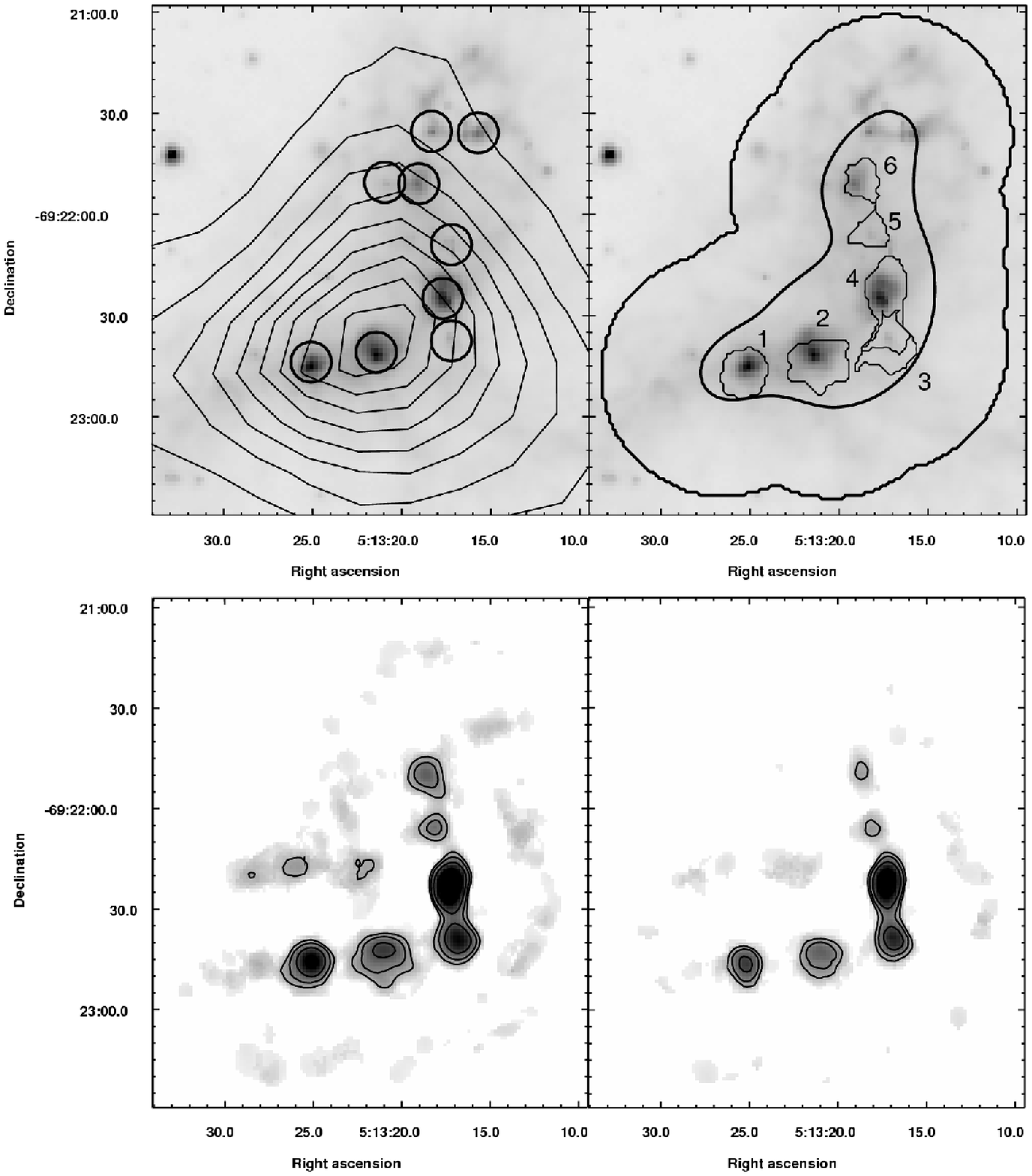}
\caption{Same as Figure 1, but for N\,113.}
\label{f2}
\end{center}
\end{figure}

\clearpage

\begin{figure}[t]
\begin{center}
\includegraphics[width=0.95\textwidth]{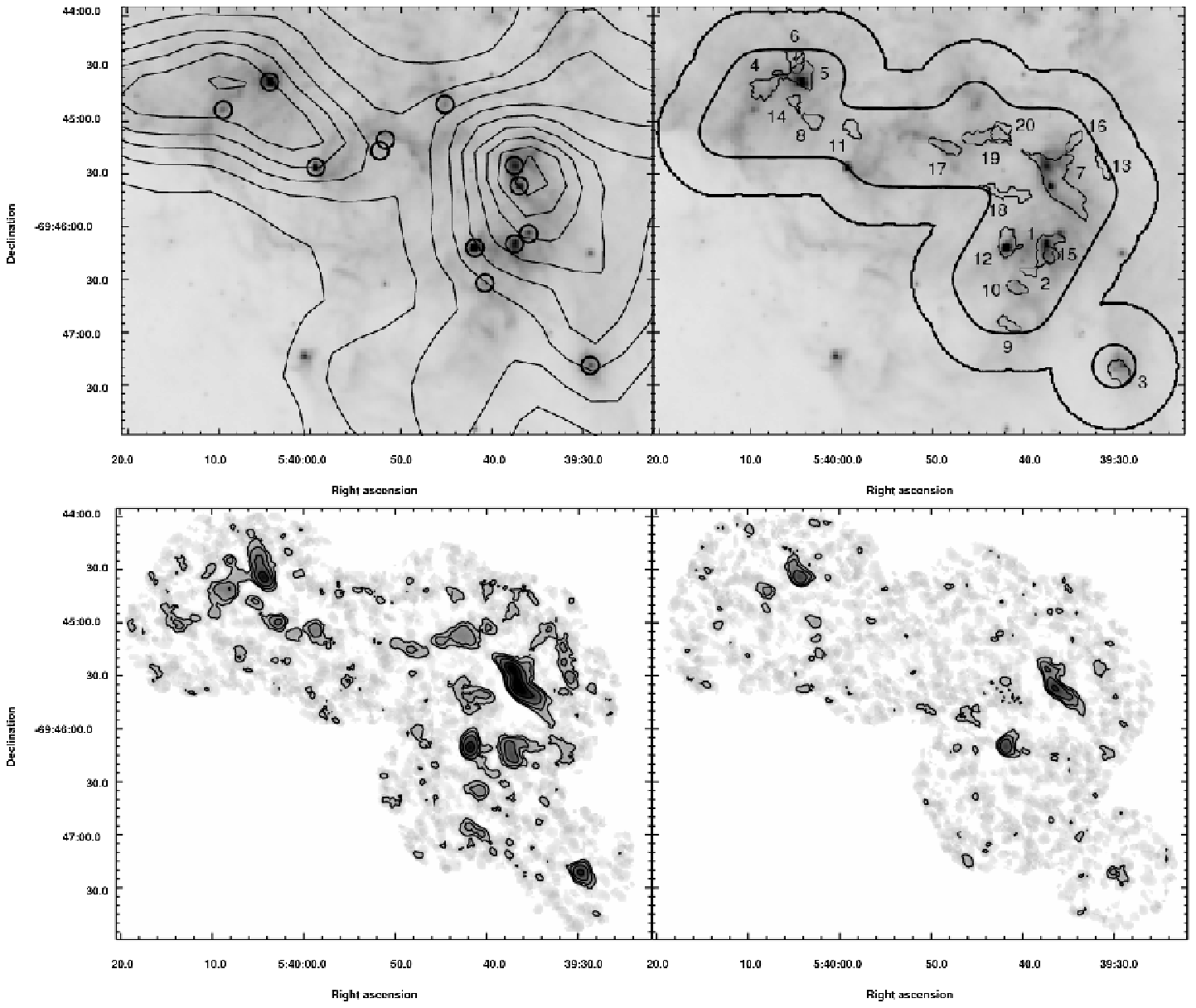}
\caption{Same as Figure 1, but for N\,159.}
\label{f3}
\end{center}
\end{figure}

\clearpage

\begin{figure}[t]
\begin{center}
\includegraphics[width=0.95\textwidth]{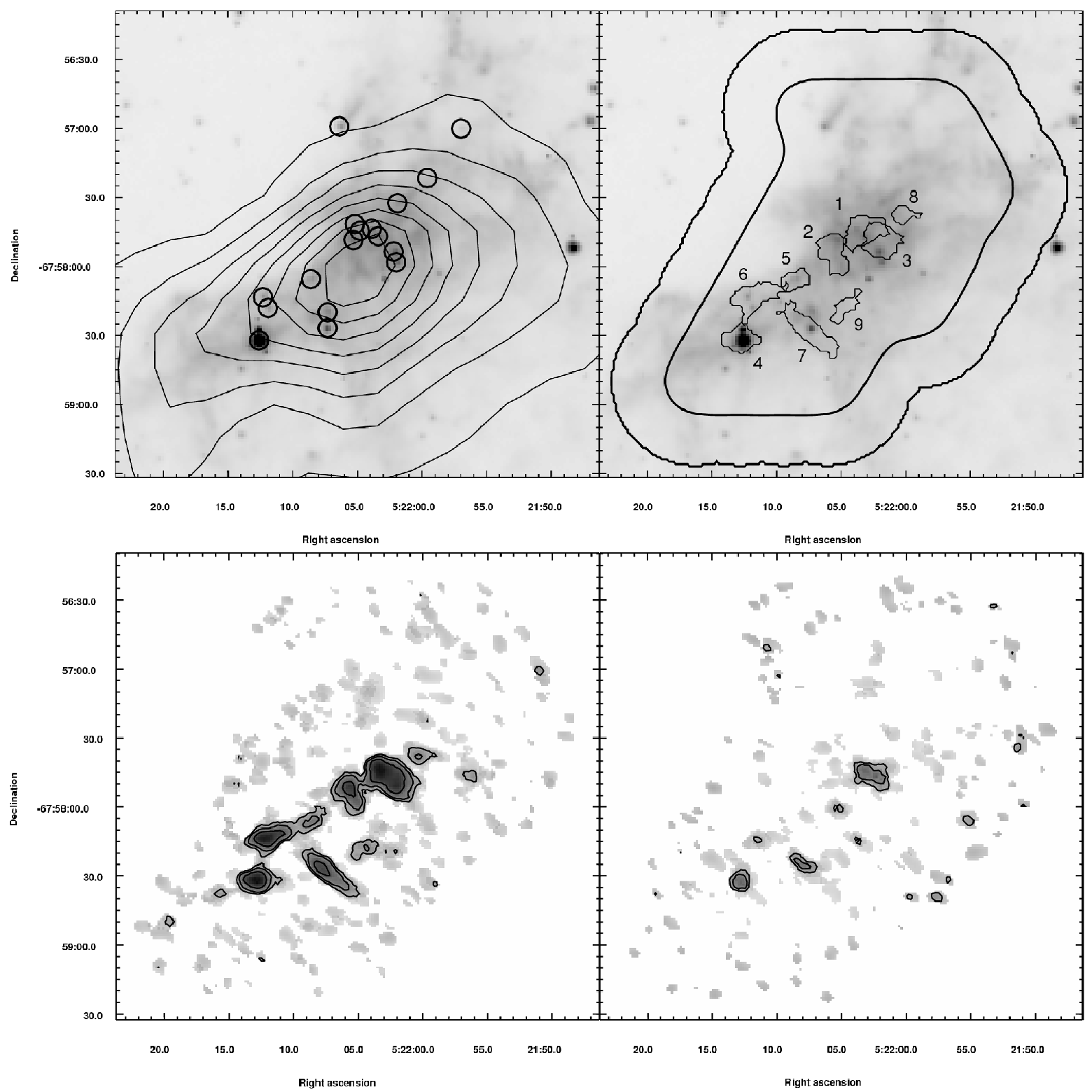}
\caption{Same as Figure 1, but for N\,44 Region 1.}
\label{f4}
\end{center}
\end{figure}

\clearpage

\begin{figure}[t]
\begin{center}
\includegraphics[width=0.95\textwidth]{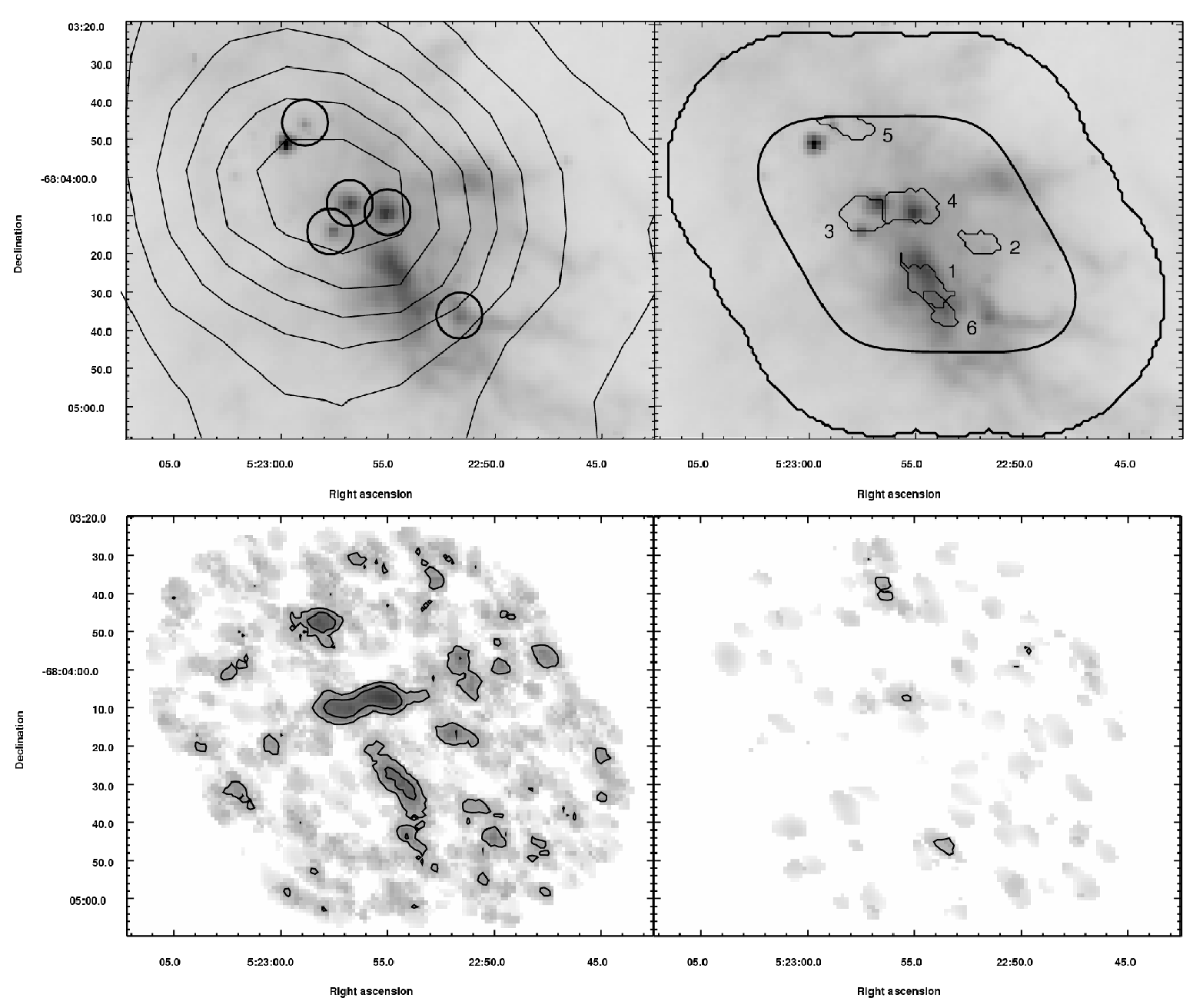}
\caption{Same as Figure 1, but for N\,44 Region 2.}
\label{f5}
\end{center}
\end{figure}

\clearpage

\begin{figure}[t]
\begin{center}
\includegraphics[width=0.8\textwidth]{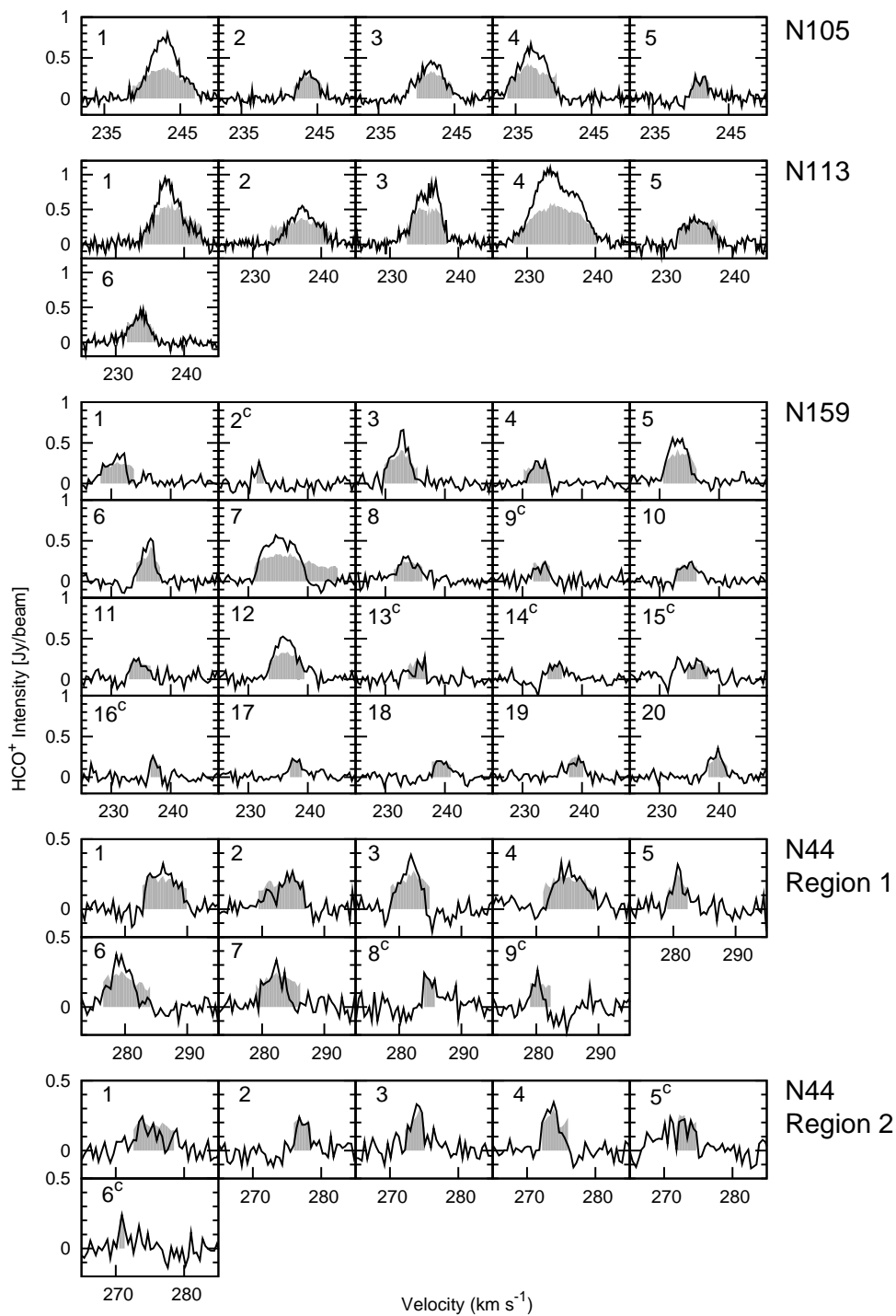}
\caption{HCO$^{+}$ spectra of the 46 clumps identified by \textsf{CPROPS}. The dark solid line shows the spectrum at the central peak of the clump's emission, while the gray shaded region displays the average spectrum for all emission associated with the clump. Candidate clumps are marked with a superscript ``c'' (see text).}
\label{f5}
\end{center}
\end{figure}

\clearpage

\begin{figure}[t]
\begin{center}
\includegraphics[width=0.3\textwidth,angle=270]{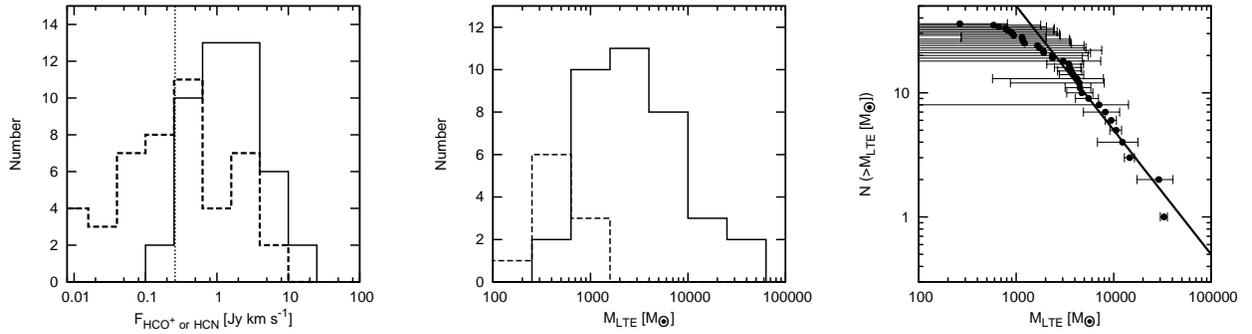}
\caption{Distribution of $F_{\rm{HCO^{+}}}$, $F_{\rm{HCN}}$, and M$_{\rm{LTE}}$ for the clump sample. The histogram displays the binned distribution, while the points show the integrated distribution. The left panel displays the distribution of $F_{\rm{HCO^{+}}}$ (solid line histogram) and $F_{\rm{HCN}}$ (dashed line histogram) for both the clumps and candidate clumps. The vertical dashed line shows the location of the sensitivity limit for a $\Delta$v = 1 km s$^{-1}$ clump in N\,113 and N\,44. In the center M$_{\rm{LTE}}$ panel the solid line histogram is for the clump population and the dashed line histogram is for the candidate clumps. The thick solid line in the right panel for the integrated M$_{\rm{LTE}}$ distribution is a power law of dN/dM$\propto$ M$^{-2}$.}
\label{f6}
\end{center}
\end{figure}

\clearpage

\begin{figure}[t]
\begin{center}
\includegraphics[width=0.7\textwidth]{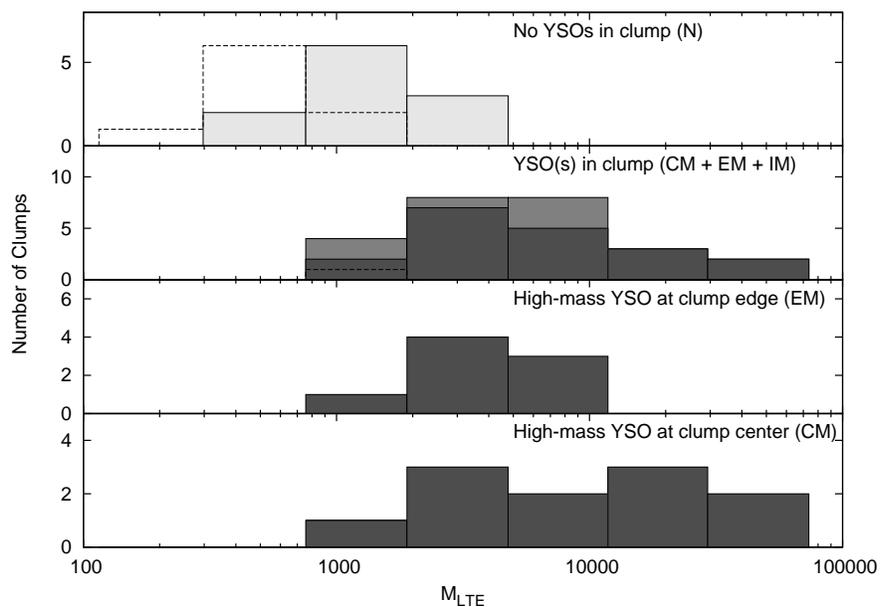}
\caption{The distribution of M$_{\rm{LTE}}$ for the clumps separated by clump groups. Top: distribution for clumps (solid line with gray fill) and candidate clumps (dotten line with no fill) of group N  ; Middle top: clumps with YSOs within their borders (groups IM, EM, and CM). Massive YSO clumps (EM and CM) are denoted by dark gray, intermediate-mass YSO clumps (IM) by a lighter gray. Middle bottom: clumps in group EM; Bottom: clumps in group CM.}
\label{f7}
\end{center}
\end{figure}

\clearpage

\begin{figure}[t]
\begin{center}
\includegraphics[width=0.7\textwidth]{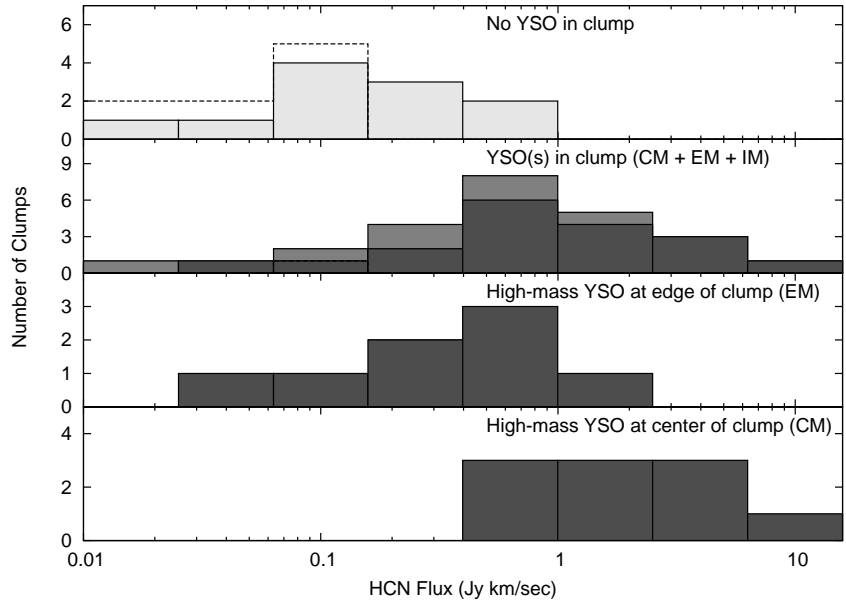}
\caption{The distribution of clump HCN fluxes separated by clump groups. Panels are as in Figure 7.}
\label{f9}
\end{center}
\end{figure}

\clearpage

\clearpage

\begin{figure}[t]
\begin{center}
\includegraphics[width=0.7\textwidth]{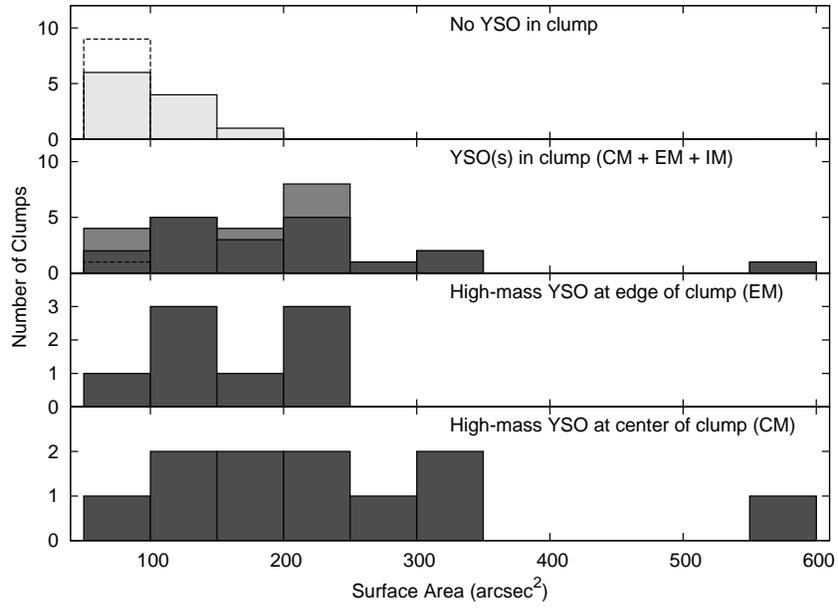}
\caption{The distribution of clump surface areas separated by clump groups. Panels are as in Figure 7.}
\label{f9}
\end{center}
\end{figure}

\clearpage

\begin{figure}[t]
\begin{center}
\includegraphics[width=0.7\textwidth]{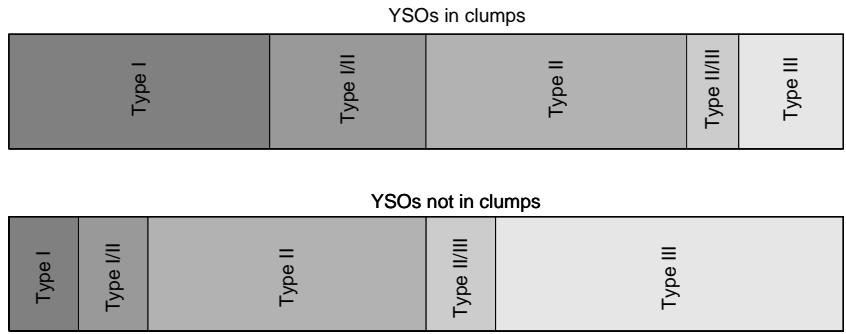}
\caption{Proportion of YSOs from \citet{chen09,chen10} of each SED type within (top panel) and outside of (bottom panel) clumps. YSO Types are indicated and differentiated by shade of gray, while the length of each Types's bar corresponds to the relative number of YSOs of that type.}
\label{f12}
\end{center}
\end{figure}

\clearpage

\begin{figure}[t]
\begin{center}
\includegraphics[width=0.7\textwidth,angle=270]{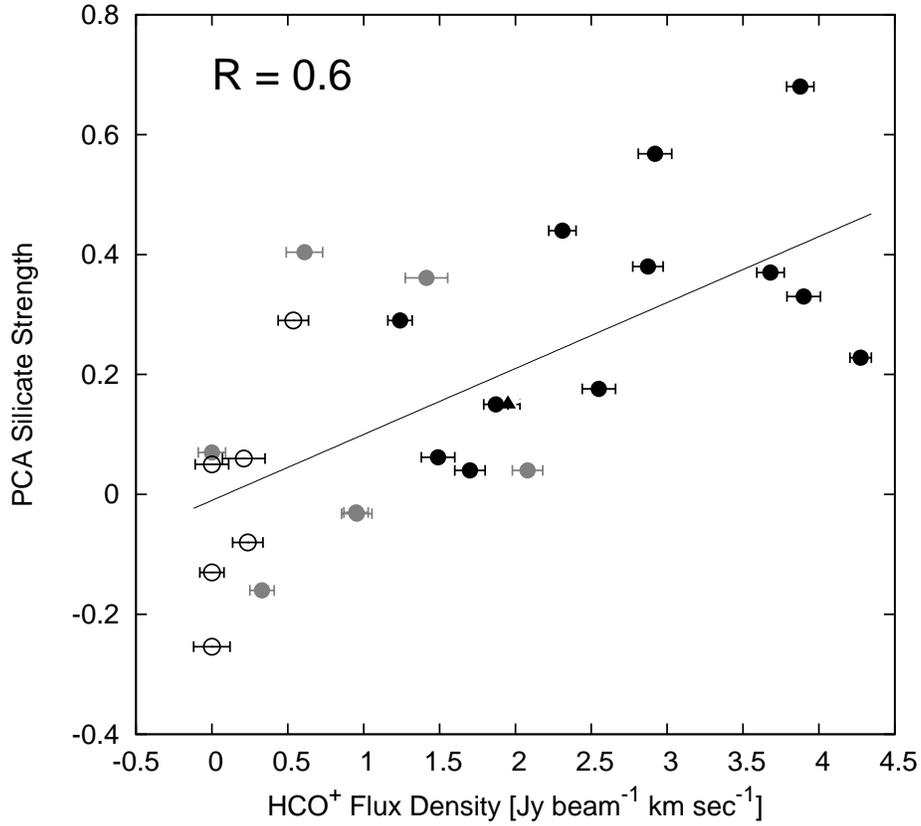}
\caption{PCA silicate strengths for the YSOs in our ATCA imaged area as a function of on-source HCO$^{+}$ flux density. Black filled circles are massive YSOs centrally located within a clump, gray filled circles are massive YSOs located on the edge of clumps, open circles are YSOs not located in clumps, and triangles are intermediate mass YSOs located within clumps. The line is a linear fit to the data to guide the eye. Spearman's rank correlation coefficient, R, is given in the upper left of the figure.}
\label{f12}
\end{center}
\end{figure}

\clearpage

\begin{figure}[t]
\begin{center}
\includegraphics[width=0.8\textwidth,angle=270]{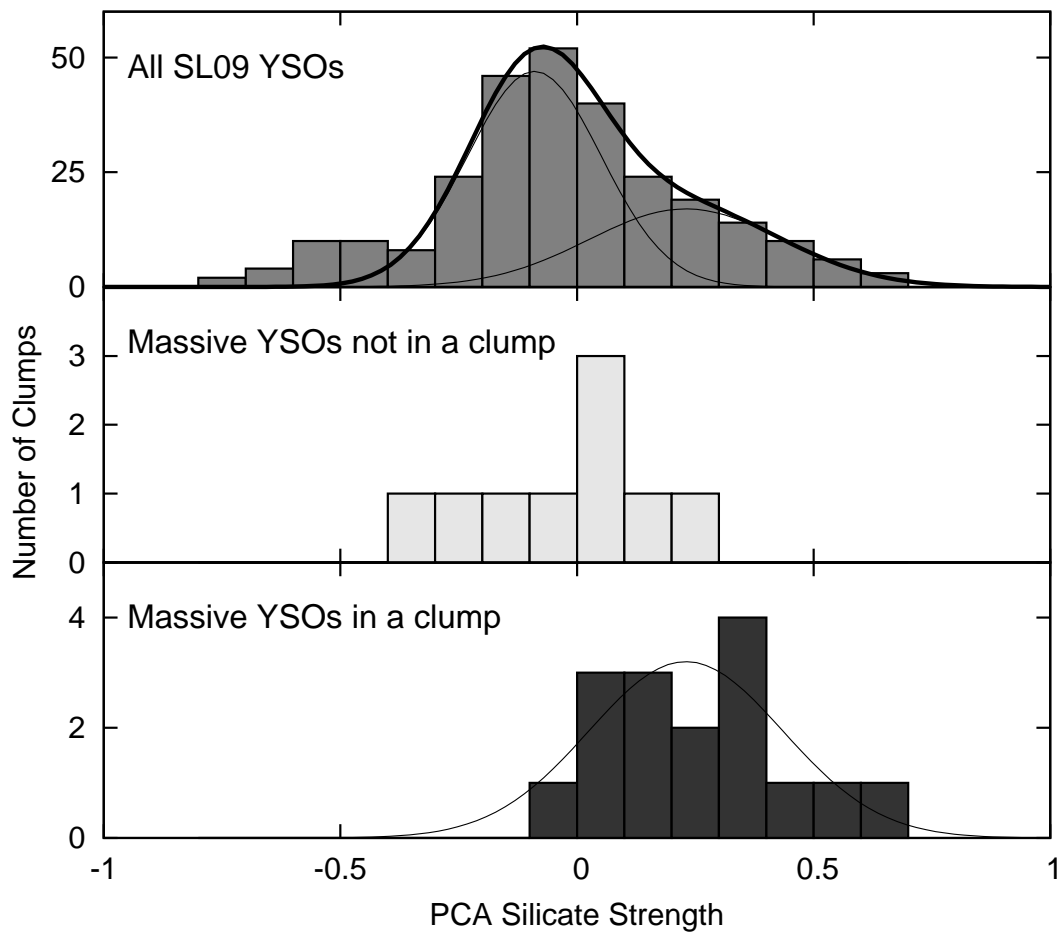}
\caption{Distribution of the strength of the 10$\mu$m silicate feature as determined by PCA for the YSOs contained within the observed areas of N\,105, N\,113, N\,159, and N\,44 (middle and bottom panels) compared to the distribution for all massive YSOs for which SL09 has \textit{Spitzer} IRS spectra (top panel). The middle panel shows the distribution for YSOs in the ATCA-images area not in clumps, while the bottom panel shows the same for YSOs located in a clump along with a Gaussian fit to the distribution. The thin curves in the top panel show the fit of this Gaussian to the SL09 catalog's high silicate strength wing (right Gaussian curve, representative of the distribution of clump-bound YSOs) and a Gaussian fit to the rest of the distribution (left Gaussian curve, representative of the distribution of YSOs outside of clumps). The dark curve is the summation of these two Gaussians.}
\label{f11}
\end{center}
\end{figure}

\clearpage

\begin{figure}[t]
\begin{center}
\includegraphics[width=0.9\textwidth]{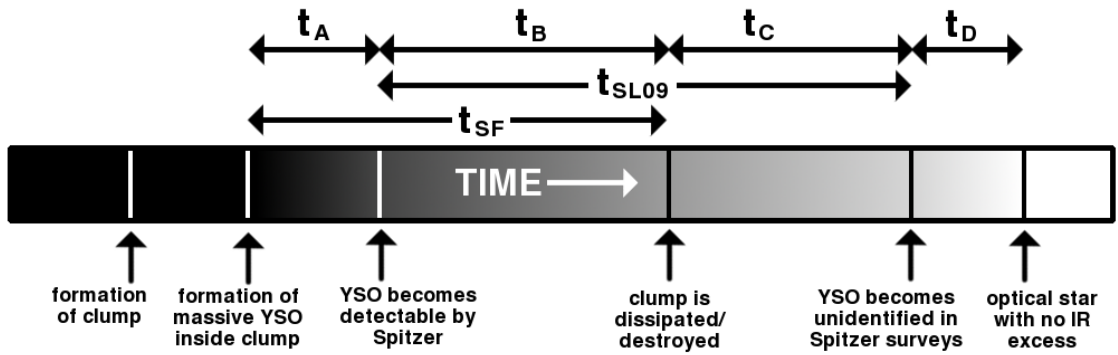}
\caption{A graphical representation of the stages of massive star forcmation within a clump. Time proceeds from left to right.}
\label{f9}
\end{center}
\end{figure}

\end{document}